\pdfoutput=1
\documentclass[letterpaper,titlepage,11pt]{article}

\usepackage{jheppub}
\usepackage[T1]{fontenc}
\usepackage[latin9]{inputenc}
\usepackage[active]{srcltx}
\usepackage{textcomp}
\usepackage{amsmath}
\usepackage{amssymb}
\usepackage{esint}
\usepackage{multirow}
\usepackage{textcomp}

\def\cG{\mathcal{G}}

\def\cL{\mathcal{L}}

\usepackage{enumerate}
\usepackage{xcolor}
\def\be{\begin{eqnarray}}
\def\ee{\end{eqnarray}}
\def\beann{\begin{eqnarray*}}
\def\eeann{\end{eqnarray*}}
\def\beq{\begin{equation}}
\def\eeq{\end{equation}}
\def\ba{\begin{array}}
\def\ea{\end{array}}
\def\ben{\begin{enumerate}}
\def\een{\end{enumerate}}
\def\bea{\begin{eqnarray}}
\def\eea{\end{eqnarray}}

\providecommand{\Lt}{{\tt L}}
\renewcommand{\Lt}{{\tt L}}

\providecommand{\Gt}{{\tt G}}
\renewcommand{\Gt}{{\tt G}}

\providecommand{\At}{{\tt A}}
\renewcommand{\At}{{\tt A}}
\providecommand{\St}{{\tt S}}
\renewcommand{\St}{{\tt S}}

\providecommand{\Tt}{{\tt T}}
\renewcommand{\Tt}{{\tt T}}

\def\bR{{\mathbb{R}}}

\def\cG{{\cal G}}

\def\cL{{\cal L}}

\def\be{\begin{equation}}
\def\ee{\end{equation}}
\def\bea{\begin{eqnarray}}
\def\eea{\end{eqnarray}}
\def\ba{\begin{array}}
\def\ea{\end{array}}
\def\nn{\nonumber}

\usepackage{amsfonts}

\usepackage{tikz}

\usepackage{bbm}

\DeclareMathOperator{\extdm}{d}
\newcommand{\extd}{\extdm \!}

\title{\bf{Exploring new Boundary Conditions for $\mathcal{N}=(1,1)$ Extended  Higher\,-\,Spin $AdS_3$ Supergravity}}
\subheader{ITU Preprint,\,\\ \today}
\newcommand{\itu}{\dagger}

\author[\itu]{H. T. \"Ozer}
\emailAdd{ozert@itu.edu.tr}
\author[\itu]{,\,\,\,Ayt\"ul Filiz}
\emailAdd{aytulfiliz@itu.edu.tr}

\affiliation[\itu]{Istanbul Technical University,\,Faculty of Science and Letters,\,Physics Department,\\34469 Maslak,\,Istanbul,Turkey.}

\abstract{In this paper,\,we present a candidate for $\mathcal{N}=(1,1)$  extended higher\,-\,spin $AdS_3$  supergravity
with the most general boundary conditions discussed by Grumiller and Riegler recently.\,We show that the
asymptotic symmetry algebra consists of two copies of the $\mathfrak{osp}(3|2)_k$ affine algebra in the
presence of the most general boundary conditions.\,Furthermore,\,we impose some certain restrictions on gauge
fields on the most general boundary conditions and that leads us  to the supersymmetric extension of the
Brown\,-\,Henneaux boundary conditions.\,We eventually see that the asymptotic symmetry algebra reduces to
two copies of the $\mathcal{SW}(\frac{3}{2},2)$ algebra for $\mathcal{N}=(1,1)$ extended higher\,-\,spin supergravity.}
\begin{document}
\maketitle
\section{Introduction}
The most common asymptotic symmetries of $AdS_3$ gravity with a negative cosmological constant in $3D$ are known as
two copies of the Virasoro algebras and this has been  written first by Brown and Henneaux in their seminal paper
\cite{Brown:1986nw}.\,So this work is well known as  both  a pioneer of $AdS_3/CFT_2$ correspondence \cite{Maldacena:1997re,Maldacena:1997re2}
and also a realization of the Holographic Principle\cite{Susskind:1994vu}.\,One of the biggest breakthroughs
in theoretical physics in the past few decades is undoubtedly the discovery of the $AdS/CFT$ correspondence
describing  an equivalence between the Einstein gravity  and a large $N$ gauge field theory.\\

The pure Einstein gravity in this context is simply a Chern\,-\,Simons gauge theory,\,that is,\,it is rewritten
as a gauge field theory,\,in such a way that the structure simplifies substantially.\,This recalls us that there
are no local propagating degrees of freedom in the theory,\,and hence no graviton in three\,-\,dimensions.\,Therefore,\,this
gauge theory is said purely topological and only global effects are of physical relevance.\,Finally,\,one
must emphasize here that the dynamics of the theory is controlled entirely by the boundary conditions,\,because
its dynamical content is far from insignificant due to the existence of boundary conditions.\,This fact was
first discovered by  Achucarro and Townsend\,\cite{Achucarro:1987vz},\,and subsequently developed by Witten
\cite{Witten:1988hc}.\,The things found here is  that the gravity action in three\,-\,dimensions and equations of
motions are in the same class with a Chern\,-\,Simons theory for a suitable gauge group.\,Under a convenient choice of
boundary conditions, there is an infinite number of degrees of freedom living on the boundary.\,These
boundary conditions are required, but these conditions are not unique in the selections.\,
The dynamical properties of the theory are also highly sensitive to the selection of these boundary conditions.\,Thus,\,the
situations of asymptotic symmetries as the residual global gauge symmetries emerge.\\

In the case of $AdS_3$ gravity above with a negative cosmological constant in $3D$,\,the most famous of these boundary
conditions is worked in the same paper \cite{Brown:1986nw}.\,These boundary conditions also contain $BTZ$ black holes
\cite{Banados:1992wn,Banados:1992gq}.\,Besides,\,the Chern\,-\,Simons higher\,-\,spin theories are  purely bosonic
theories \cite{Bergshoeff:1989ns,Blencowe:1988gj} as versions of Vasiliev higher\,-\,spin theories
\cite{Vasiliev:1999ba,Vasiliev:2000rn} with higher\,-\,spin fields of integer spin,\,they are based on the
$\mathfrak{sl}(N ,\bR)$ algebras  and higher\,-\,spin algebras $\mathfrak{hs}(\lambda)$  respectively.\,
The Chern\,-\,Simons higher\,-\,spin theories can also be resulted with a realization of  the  classical $W_N$ asymptotic symmetry
algebras as in  the related  two\,-\,dimensional CFT's
\cite{Henneaux:2010xg,Campoleoni:2010zq,Campoleoni:2011hg,Campoleoni:2018uib,Tan:2011tj,Ozer:2017awd}.\,The
validity of these results can be extended to supergravity theory \cite{Brown:1986nw,Tan:2012xi,Banados:1998pi}, as well as a higher\,-\,spin theory
\cite{Blencowe:1988gj,Bergshoeff:1989ns}.\,Beyond that a supersymmetric generalization of these bosonic theories can be
achieved by considering Chern\,-\,Simons theories based on superalgebras such as  $\mathfrak{sl}(N|N-1)$,\,see e.g
\cite{Candu:2012jq,Henneaux:2012ny,Tan:2012xi,Banados:1998pi,Peng:2012ae,Hanaki:2012yf},\,or $\mathfrak{osp}(N|N-1)$
\cite{Chen:2013oxa} which can be obtained  by truncating out all the odd spin generators and one copy of the fermionic
operators in  $\mathfrak{sl}(N|N-1)$.\\

The critical role of boundary conditions in field theories and in particular in gravity theories has been well understood
so far.\,In three dimensions,\,as previously mentioned,\,asymptotic $AdS_3$ boundary conditions by Brown and Henneaux \cite{Brown:1986nw}
provided an important precursor to the $AdS_3/CFT_2$.\,For a long time,\,these boundary conditions have been changed
(see, eg \cite{Grumiller:2016pqb,Afshar:2015wjm,Compere:2013bya,Afshar:2016wfy,Troessaert:2013fma,Avery:2013dja,Perez:2016vqo}) and generalized (see, eg \cite{Henneaux:2002wm,Henneaux:2004zi,Grumiller:2008es,Henneaux:2009pw,Oliva:2009ip,Barnich:2013sxa,Bunster:2014cna,Perez:2015jxn})
many times.\,Our motivation is to construct a candidate solution for the most general $\mathcal{N}=(1,1)$ extended higher\,-\,spin
supergravity theory in $AdS_3$.\,Our theory falls under the same metric class as the recently constructed most general
$AdS_3$ boundary condition by Grumiller and Riegler\cite{Grumiller:2016pqb}.\,They showed that all charges and chemical
potentials that appear in the Chern-Simons formulation can also be seen in the metric formulation.\,It is also mentioned that the possibility
to obtain asymptotic symmetries that are the affine version of the gauge algebras of the Chern-Simons theory
is not a surprise given the relation between Chern-Simons theories and Wess-Zumino-Witten models\,\cite{Elitzur:1989nr}.\,If one examines
the correspondence between 2+1 dimensional pure Chern-Simons gauge theories  and 1+1 dimensional Wess-Zumino-Witten
conformal field theories,\,it can be seen that this  is the core of the relationship between these theories.
\,This method recently has also been applied to flat-space \cite{Grumiller:2017sjh} and chiral higher\,-\,spin gravity \cite{Krishnan:2017xct} which
showed a new class of boundary conditions for higher\,-\,spin theories in $AdS_3$.\,The simplest extension of the
Grumiller and Riegler's analysis for the most general $\mathcal{N}=(1,1)$ and $\mathcal{N}=(2,2)$ extended higher
spin supergravity theory in $AdS_3$ is the Valc$\acute{a}$rcel's paper \cite{Valcarcel:2018kwd}
where the asymptotic symmetry algebra for the loosest set of boundary conditions for (extended) supergravity has
been obtained.\,This is an alternative to the non\,-\,chiral Drinfeld\,-\,Sokolov type boundary conditions.\,In particular,\,
we first focus on  the simplest example,\,$\mathcal{N}=(1,1)$ Chern\,-\,Simons theory based on the superalgebra $\mathfrak{osp}(1|2)$.\,The
related asymptotic symmetry algebra is two copies of the $\mathfrak{osp}(1|2)_k$ affine algebra.\,Then
one can extend this study to the $\mathcal{N}=(1,1)$ Chern\,-\,Simons theory based on the superalgebra
$\mathfrak{osp}(3|2)$ and the symmetry algebras are given by  two copies of the $\mathfrak{osp}(3|2)_k$ affine
algebra.\,Furthermore,\,we also impose certain restrictions on the gauge fields on the most general boundary
conditions and that leads to the supersymmetric extensions of the Brown\,-\,Henneaux boundary conditions.
\,From these restrictions,\,we see that the  asymptotic symmetry algebras reduce to two copies of the
$\mathcal{SW}(\frac{3}{2},2)$ algebra for the most general $\mathcal{N}=(1,1)$ extended higher\,-\,spin
supergravity theory in $AdS_3$.\,It will be also very interesting to show  another class of boundary conditions
that appeared in the literature (see,\,eg \cite{Compere:2013bya,Afshar:2016wfy,Troessaert:2013fma,Avery:2013dja})
for (super)gravity case,\,whose higher\,-\,spin generalization is less clear than the Grumiller\,-\,Riegler boundary conditions.
\,Therefore,\,one can think that this method provides a good laboratory for investigating the rich asymptotic structure of
extended supergravity.\\

The outline of the paper is as follows.\,In the next section,\,we give briefly a fundamental formulation of
$\mathcal{N}=(1,1)$  supergravity in the perspective of $\mathfrak{osp}(1|2) \oplus \mathfrak{osp}(1|2)$
Chern\,-\,Simons gauge theory for both affine and superconformal boundaries respectively in three-dimensions.\,
In section 3,\,we carry out our calculations to extend the theory to $\mathfrak{osp}(3|2) \oplus \mathfrak{osp}(3|2)$
higher\,-\,spin Chern\,-\,Simons supergravity in the presence of both affine and superconformal boundaries,\,in where
we showed explicitly principal embedding of  $\mathfrak{osp}(1|2) \oplus \mathfrak{osp}(1|2)$ and also demonstrated
how asymptotic symmetry and higher\,-\,spin Ward identities arise from the bulk equations of motion coupled to spin
$s$,\,($s\,=\,\frac{3}{2},2,2,\frac{5}{2}$) currents.\,This section is devoted to the case of classical two copies of the
$\mathfrak{osp}(3|2)_k$ affine algebra on the affine boundary and $\mathcal{SW}(\frac{3}{2},2)$  symmetry algebra on the
superconformal boundary  as asymptotic symmetry algebras,\,and also the chemical potentials related to source fields
appearing through the temporal components of the connection are obtained.\,In section 4,\,we present two most general
boundary conditions for pure bosonic gravity and $\mathcal{N}=1$ extended supergravity.\,That is,
\,The Avery\,-\,Poojary\,-\,Suryanarayana gravity for $\mathfrak{sl}(2,\mathbb{R})\oplus\mathfrak{sl}(2)$ and
$\mathfrak{osp}(1|2)\oplus\mathfrak{sl}(2)$ respectively with some further checks.\,On the other hand,\,it is shown
that  the Chern\,-\,Simons action which compatible with our boundary conditions leads to a finite action and
a well\,-\,defined variational principle  for the higher\,-\,spin fields.\,The final section contains our summary and conclusion.


\section{Supergravity in Three\,-\,Dimensions, A Review:}\label{sec2}
In this section,\,we review the Chern\,-\,Simons formalism for higher\,-\,spin supergravity.\,In particular,\,we use
this formalism to study supergravity in the $\mathfrak{osp}(1|2)$ superalgebra basis under the
same metric class as the recently constructed most general  $AdS_3$ boundary condition by Grumiller and Riegler
\cite{Grumiller:2016pqb}.\,
\subsection{Connection to Chern\,-\,Simons Theory}
The three\,-\,dimensional Einstein\,-\,Hilbert action for $\mathcal{N}=(1,1)$ supergravity,
\,which is defined on any manifold $\mathcal{M}$,\,with a negative cosmological constant
is classically equivalent to the Chern\,-\,Simons action, \,as it was first proposed by Achucarro and Townsend in
\cite{Achucarro:1987vz} and developed by Witten in \cite{Witten:1988hc}.\,One can start by defining  1\,-\,forms
$(\Gamma,\bar{\Gamma})$ taking values in the gauge group's $\mathfrak{osp}(1|2)$ superalgebra, and also the
supertrace $\mathfrak{str}$ is taken over the superalgebra generators.\,The Chern\,-\,Simons action can be written in the form,
\begin{equation}
S = S_{\textrm{\tiny CS}}[\Gamma] - S_{\textrm{\tiny CS}}[\bar{\Gamma}]\\
\end{equation}
where
\begin{equation}
S_{\textrm{\tiny CS}}[\Gamma] = \frac{k}{4\pi}\int_{\mathcal{M}} \langle\Gamma\wedge \mathrm{d}\Gamma\,+\,\frac{2}{3}\,\Gamma\wedge\Gamma\wedge\Gamma \rangle\label{CSA}.
\end{equation}
Here $k=\frac{\ell}{4G\mathfrak{str}( {\tt L}_{0} {\tt L}_{0}) }=\frac{c}{6\mathfrak{str}( {\tt L}_{0} {\tt L}_{0}) }$ is the level
 of the Chern\,-\,Simons theory depending on the $AdS$ radius $l$ and Newton's constant $G$ with the related  central charge $c$ of
 the superconformal field theory.
 \,If $\Lt_{i},(i=\pm1,0)$ and $\Gt_{p},(p=\pm\frac{1}{2})$  are the generators of $\mathfrak{osp}(1|2)$ superalgebra.
\,We have expressed $\mathfrak{osp}(1|2)$ superalgebra such that
\be
\label{algebraLG}
\left[\Lt_{i},\Lt_{j}\right]= \left(i-j\right)\Lt_{i+j},\;\qquad\left[\Lt_{i},\Gt_{p}\right]=\big(\frac{i}{2}-p\big)\Gt_{i+p},\;\qquad\left\{\Gt_{p},\Gt_{q}\right\} =-2\,\Lt_{p+q}.
\ee
The equations of motion for the Chern\,-\,Simons gauge theory give the flatness condition $F = \bar{F} =0$ where
\begin{equation}
\label{flatness}
F= \mathrm{d}\Gamma + \Gamma\wedge \Gamma =0
\end{equation}
is the same as Einstein's equation.
$\Gamma$ and $\bar{\Gamma}$ are related to the metric $g_{\mu \nu}$ through the veilbein $e = \frac{\ell}{2}(\Gamma- \bar{\Gamma})$
\begin{equation}\label{metric}
g_{\mu \nu} =  \frac{1}{2}\,\mathfrak{str}( e_{\mu} e_{\nu} ).
\end{equation}
One can choose a radial gauge of the form
\begin{eqnarray}
\label{ads31}
\Gamma&=& b^{-1} a\left(t,\phi\right) b +b^{-1} db,\,\,\, \bar{\Gamma}=b\bar{a}\left(t,\phi\right)b^{-1} + bdb^{-1}
\end{eqnarray}
with state-independent group element as\cite{Grumiller:2016pqb}
\begin{eqnarray}\label{gr}
b(\rho)=e^{\Lt_{-1}}e^{\rho \Lt_{0}}
\end{eqnarray}
which manifests all the $\mathfrak{osp}(1|2)$ charges and chemical
 potentials and also the choice of $b$ is irrelevant in the case of asymptotic symmetry, as long as $\delta b=0$.\,Moreover,
 \,$a\left(t,\phi\right)$ and $\bar{a}\left(t,\phi\right)$  in the radial gauge are the $\mathfrak{osp}(1|2)$ superalgebra
 valued fields,\,which are independent from the radial coordinate,\,$\rho$ as
\begin{eqnarray}\label{conn}
a\left(t,\varphi\right) = a_{t}\left(t,\varphi\right)\mathrm{d}t+a_{\varphi}\left(t,\varphi\right)\mathrm{d}\varphi.
\end{eqnarray}
\subsection{$\mathfrak{osp}(1|2) \oplus \mathfrak{osp}(1|2)$ Chern\,-\,Simons $\mathcal{N}=(1,1)$ Supergravity  for Affine Boundary}\label{osp21}
The affine case is given by reviewing asymptotically $AdS_3$ boundary conditions  for a $\mathfrak{osp}(1|2) \oplus \mathfrak{osp}(1|2)$ Chern\,-\,Simons theory, and how to determine the asymptotic symmetry algebra using the method described in \cite{Grumiller:2016pqb}.\,Thus the most general solution of Einstein's equation  that is asymptotically $AdS_3$,\,as a generalization of Fefferman\,-\,Graham method is given by with a flat boundary metric
\begin{eqnarray}\label{sugra08}
    \mathrm{d}s^2 &=& \mathrm{d}\rho^2 + 2\left[ e^\rho N^{(0)}_i +  N^{(1)}_i +  e^{-\rho} N^{(2)}_i + \mathcal O \left( e^{-2\rho}\right)\right]\mathrm{d}\rho \mathrm{d}x^i\nonumber\\
    &+& \left[ e^{2\rho} g^{(0)}_{ij} +  e^\rho g^{(1)}_{ij} +  g^{(2)}_{ij} + \mathcal O \left( e^{-\rho}\right)\right]\mathrm{d}x^i \mathrm{d}x^j.
\end{eqnarray}
We need to choose the most general boundary conditions for $\mathcal{N}=(1,1)$ supergravity such that they maintain this metric form.\,In the following,\,we only focus on the $\Gamma$\,-\,sector.\,Therefore,\,one can propose to write the components of the $\mathfrak{osp}(1|2)$ superalgebra valued connection in the form,
\begin{eqnarray}\label{bouncond999}
a_\varphi\left(t,\varphi\right) &=&
\sum_{i=-1}^{+1}\alpha_i\mathcal{L}^i\left(t,\varphi\right) \Lt_{i}
+\sum_{p=-1/2}^{+1/2}\beta_p\mathcal{G}^{p}\left(t,\varphi\right)\Gt_{p}
\end{eqnarray}
where $\left(\alpha_i,\beta_p\right)$'s are some scaling parameters to be determined later and we have five functions: three bosonic $\mathcal{L}^i$ and two fermionic $\mathcal{G}^{p}$.\,They are usually called $charges$ and also the time component of the connection $a\left(t,\varphi\right)$
\begin{eqnarray}\label{bouncond888}
a_t\left(t,\varphi\right) &=&
\sum_{i=-1}^{+1}\mathcal{\mu}^i\left(t,\varphi\right) \Lt_{i}
+\sum_{p=-1/2}^{+1/2}\mathcal{\nu}^{p}\left(t,\varphi\right)\Gt_{p}.
\end{eqnarray}
Here,\,the time component has in total of five independent functions $(\mathcal{\mu}^i,\mathcal{\nu}^{p})$.\,They are usually called $chemical~~potentials$.\,But,\,they are not allowed to vary
\begin{eqnarray}\label{bouncond777}
\delta a_\varphi &=&
\sum_{i=-1}^{+1}\alpha_i\delta \mathcal{L}^i \Lt_{i}
+\sum_{p=-1/2}^{+1/2}\beta_p\delta \mathcal{G}^{p}\Gt_{p},\\
\delta a_t&=&0.
\end{eqnarray}
The flat connection conditions $(\ref{flatness})$ for fixed chemical potentials impose the following additional conditions as the temporal evolution of the five independent source fields $(\mathcal{L}^i,\mathcal{G}^{p})$ as
\begin{eqnarray}
\alpha_{\pm 1}\partial_{t}\mathcal{L}^{\pm 1}	&=&	
\partial_{\varphi}\mu^{\pm 1}
\mp \alpha_0\mathcal{L}^{0}\mu^{\pm 1}
\pm \alpha_{\pm 1} \mathcal{L}^{\pm 1}\mu^{0}
-2 \beta_{\pm\frac{1}{2}}\mathcal{G}^{\pm\frac{1}{2}}\nu^{\pm\frac{1}{2}},\label{eva1}\\
\frac{1}{2}\alpha_0\partial_{t}\mathcal{L}^{0}	&=&	
\frac{1}{2} \partial_{\varphi}\mu^{0}
+ \alpha_{+1}\mathcal{L}^{+1}\mu^{-1}
- \alpha_{-1}\mathcal{L}^{-1}\mu^{+1}
-\beta_{+\frac{1}{2}}\mathcal{G}^{+\frac{1}{2}}\nu^{-\frac{1}{2}}
-\beta_{-\frac{1}{2}}\mathcal{G}^{-\frac{1}{2}}\nu^{+\frac{1}{2}},\label{eva2}\\
\beta_{\pm\frac{1}{2}}\partial_{t}\mathcal{G}^{\pm\frac{1}{2}} &=&
\partial_{\varphi}\nu^{\pm\frac{1}{2}}
\pm \alpha_{\pm 1}\mathcal{L}^{\pm}\nu^{\mp\frac{1}{2}}
\mp \frac{1}{2} \alpha_0 \mathcal{L}^{0}\nu^{\pm\frac{1}{2}}
\mp \beta_{\mp \frac{1}{2}}\mu^{\pm}\mathcal{G}^{\mp\frac{1}{2}}
\pm\frac{1}{2}\beta_{\pm \frac{1}{2}}\mu^{0}\mathcal{G}^{\pm\frac{1}{2}}.\label{eva3}
\end{eqnarray}
After the temporal evolution of the source fields,\,one can start to compute the gauge transformations for asymptotic symmetry algebra by considering all transformations
\begin{equation}
    \delta_{\lambda}\Gamma = \mathrm{d}\lambda + \left[\Gamma,\lambda \right] \label{boundarycond}
\end{equation}
that preserve the boundary conditions with the gauge parameter in the $\mathfrak{osp}(1|2)$ superalgebra
 \begin{equation}
    \lambda=b^{-1}\bigg[
    \sum_{i=-1}^{+1}\mathcal{\epsilon}^i\left(t,\varphi\right) \Lt_{i}
+\sum_{p=-1/2}^{+1/2}\mathcal{\zeta}^{p}\left(t,\varphi\right)\Gt_{p}
    \bigg]b.\label{boundarycond2}
\end{equation}
Here,\,the gauge parameter has in total of five arbitrary  functions : bosonic $\mathcal{\epsilon}^i$ and fermionic $\mathcal{\zeta}^{p}$ on the boundary.\,
The condition $(\ref{boundarycond})$ impose that transformations on the gauge are given by
\begin{eqnarray}\label{evaa11}
\alpha_{\pm 1}\delta_{\lambda}\mathcal{L}^{\pm 1}	&=&	
\partial_{\varphi}\epsilon^{\pm 1}
\mp \alpha_0\mathcal{L}^{0}\epsilon^{\pm 1}
\pm \alpha_{\pm 1} \mathcal{L}^{\pm 1}\epsilon^{0}
-2 \beta_{\pm\frac{1}{2}}\mathcal{G}^{\pm\frac{1}{2}}\zeta^{\pm\frac{1}{2}},\label{evaa21}\\
\frac{1}{2}\alpha_0\delta_{\lambda}\mathcal{L}^{0}	&=&	
\frac{1}{2} \partial_{\varphi}\epsilon^{0}
+ \alpha_{+1}\mathcal{L}^{+1}\epsilon^{-1}
- \alpha_{-1}\mathcal{L}^{-1}\epsilon^{+1}
-\beta_{+\frac{1}{2}}\mathcal{G}^{+\frac{1}{2}}\zeta^{-\frac{1}{2}}
-\beta_{-\frac{1}{2}}\mathcal{G}^{-\frac{1}{2}}\zeta^{+\frac{1}{2}},\\
\beta_{\pm\frac{1}{2}}\delta_{\lambda}\mathcal{G}^{\pm\frac{1}{2}} &=&
\partial_{\varphi}\zeta^{\pm\frac{1}{2}}
\pm \alpha_{\pm 1}\mathcal{L}^{\pm}\zeta^{\mp\frac{1}{2}}
\mp \frac{1}{2} \alpha_0 \mathcal{L}^{0}\zeta^{\pm\frac{1}{2}}
\mp \beta_{\mp \frac{1}{2}}\epsilon^{\pm}\mathcal{G}^{\mp\frac{1}{2}}
\pm\frac{1}{2}\beta_{\pm \frac{1}{2}}\epsilon^{0}\mathcal{G}^{\pm\frac{1}{2}}\label{evaa31}.
\end{eqnarray}
With analogical reasoning,\,they obey the following transformations
\begin{eqnarray}
\alpha_{\pm 1}\delta_{\lambda}\mu^{\pm 1}	&=&	
\partial_{t}\epsilon^{\pm 1}
\mp \alpha_0\mu^{0}\epsilon^{\pm 1}
\pm \alpha_{\pm 1} \mu^{\pm 1}\epsilon^{0}
-2 \beta_{\pm\frac{1}{2}}\nu^{\pm\frac{1}{2}}\zeta^{\pm\frac{1}{2}},\label{tra11}\\
\frac{1}{2}\alpha_0\delta_{\lambda}\mu^{0}	&=&	
\frac{1}{2} \partial_{t}\epsilon^{0}
+ \alpha_{+1}\mu^{+1}\epsilon^{-1}
- \alpha_{-1}\mu^{-1}\epsilon^{+1}
-\beta_{+\frac{1}{2}}\nu^{+\frac{1}{2}}\zeta^{-\frac{1}{2}}
-\beta_{-\frac{1}{2}}\nu^{-\frac{1}{2}}\zeta^{+\frac{1}{2}},\label{tra12}\\
\beta_{\pm\frac{1}{2}}\delta_{\lambda}\nu^{\pm\frac{1}{2}} &=&
\partial_{t}\zeta^{\pm\frac{1}{2}}
\pm \alpha_{\pm 1}\mu^{\pm}\zeta^{\mp\frac{1}{2}}
\mp \frac{1}{2} \alpha_0 \mu^{0}\zeta^{\pm\frac{1}{2}}
\mp \beta_{\mp \frac{1}{2}}\epsilon^{\pm}\nu^{\mp\frac{1}{2}}
\pm\frac{1}{2}\beta_{\pm \frac{1}{2}}\epsilon^{0}\nu^{\pm\frac{1}{2}}.\label{tra13}
\end{eqnarray}
As a final step,\,one now has to determine the canonical boundary charge $\mathcal{Q[\lambda]}$ that generates
the transformations $(\ref{evaa11})$-$(\ref{evaa31})$.\,Therefore, the corresponding variation of the boundary charge $\mathcal{Q[\lambda]}$ \cite{Banados:1998gg,Carlip:2005zn,Banados:1994tn,Banados:1998ta},
\,to show the asymptotic symmetry algebra,\,is given by
\begin{equation}\label{Qvar}
\delta_\lambda \mathcal{Q}=\frac{k}{2\pi}\int\mathrm{d}\varphi\;\mathfrak{str}\left(\lambda \delta \Gamma_{\varphi}\right).
\end{equation}
The canonical boundary charge $\mathcal{Q[\lambda]}$  can be integrated which reads
\begin{equation}\label{boundaryco10}
\mathcal{Q[\lambda]}=\int\mathrm{d}\varphi\;\left[\mathcal{L}^{0}\epsilon^{0}
+\mathcal{L}^{+1}\epsilon^{-1}+\mathcal{L}^{-1}\epsilon^{+1}+\mathcal{G}^{+\frac{1}{2}}
\zeta^{-\frac{1}{2}}+\mathcal{G}^{-\frac{1}{2}}\zeta^{+\frac{1}{2}}\right].
\end{equation}
We now prefer to work in complex coordinates for the affine boundary,\,$z(\bar{z})\equiv \varphi \pm i\,\frac{t}{\ell}$.\,After both the infinitesimal transformations  and the canonical boundary charge  have been determined, one can yield the Poisson bracket algebra  by using the methods \cite{Blgojevic:2002} with
\begin{equation}\label{Qvar2}
\delta _{\lambda}\digamma\,=\,\{\digamma,\mathcal{Q}[\lambda]\}
\end{equation}
for any phase space functional $\digamma$ :
\begin{eqnarray}\label{bracketalg}
\{\mathcal{L}^i(z_1),\mathcal{L}^j(z_2)\}_{_{PB}}\,&=&\,(i-j) \mathcal{L}^{i+j}(z_2)\delta(z_1-z_2)-\frac{1}{\alpha_{+1}}\eta^{ij}\partial_{\varphi}\delta(z_1-z_2),\\
\{\mathcal{L}^i(z_1),\mathcal{G}^p(z_2)\}_{_{PB}}\,&=&\,\left(\frac{i}{2}-p\right) \mathcal{G}^{i+p}(z_2)\delta(z_1-z_2),\\
\{\mathcal{G}^p(z_1),\mathcal{G}^q(z_2)\}_{_{PB}}\,&=&\,-2\mathcal{L}^{p+q}(z_2)\delta(z_1-z_2)-\frac{1}{\alpha_{+1}}\eta^{pq}\partial_{\varphi}\delta(z_1-z_2)
\end{eqnarray}
where $\alpha_{+1}=-\frac{2 \pi}{k}$ for the convention in the literature and $\eta^{ij}$ is the bilinear form in the fundamental representation of $\mathfrak{osp}(1|2)$ superalgebra.\,One can also expand $\mathcal{L}(z)$ and $\mathcal{G}(z)$ charges into Fourier modes\,$\mathcal{L}^{i}(z)=\frac{1}{2\pi}\sum_n{\Lt_{n}^{i}}z^{-n-1}$,\,and \,$\mathcal{G}^{p}(z)=\frac{1}{2\pi}\sum_r{\Gt_{r}^{p}}z^{-r-\frac{1}{2}}$,\,and also replacing $i\{\cdot,\cdot\}_{_{PB}}\rightarrow[\cdot,\cdot]$.\,A mode algebra can then be defined as:
\begin{eqnarray}
\left[\Lt_{m}^{i},\Lt_{n}^{j}\right] &=& \left(i-j\right)\Lt_{m+n}^{i+j}+k n\eta^{ij}\delta_{m+n,0}\label{com1},\\
\left[\Lt_{m}^{i},\Gt_{r}^{p}\right] &=& \left(\frac{i}{2}-p\right)\Gt_{m+r}^{i+p}\label{com2},\\
\left\{\Gt_{r}^{p},\Gt_{s}^{q}\right\} &=& -2\Lt_{r+s}^{p+q}+ k s\kappa^{pq}\delta_{r+s,0}.\label{acom1}
\end{eqnarray}
Besides,\,this mode algebra in this space is equivalent to operator product algebra,
\begin{eqnarray}\label{ope}
\mathcal{L}^{i}(z_1)\mathcal{L}^{j}(z_2)\,& \sim &\, \frac{\frac{k}{2}\eta^{ij}}{z_{12}^{2}}\,+\,\frac{(i-j)}{z_{12}} \mathcal{L}^{i+j}(z_2),\\
\mathcal{L}^{i}(z_1)\mathcal{G}^{p}(z_2)\,& \sim &\,\frac{({i\over 2}-p)}{z_{12}} \mathcal{G}^{i+p}(z_2),\\
\mathcal{G}^{p}(z_1)\mathcal{G}^{q}(z_2)\,& \sim &\, \frac{\frac{k}{2}\eta^{pq}}{z_{12}^{2}}\,+\,\frac{2}{z_{12}} \mathcal{L}^{p+q}(z_2)
\end{eqnarray}
where $z_{12}=z_1-z_2$,\,or in the more compact form,
\begin{eqnarray}\label{ope11}
\mathfrak{\mathcal{\mathfrak{J}}}^{A}(z_1)\mathfrak{\mathcal{\mathfrak{J}}}^{B}(z_2)\,& \sim &\, \frac{\frac{k}{2}\eta^{AB}}{z_{12}^{2}}\,+\,\frac{\mathfrak{\mathcal{\mathfrak{f}}}^{AB}_{~~~C} \mathfrak{\mathcal{\mathfrak{J}}}^{C}(z_2)}{z_{12}}.
\end{eqnarray}
Here, $\eta^{AB}$ is the supertrace matrix and $\mathfrak{\mathcal{\mathfrak{f}}}^{AB}_{~~C}$'s are the structure constants of the related algebra with $(A,B=0,\pm1,\pm\frac{1}{2})$,\,i.e,\,$\eta^{ip}=0$ and $\mathfrak{\mathcal{\mathfrak{f}}}^{ij}_{~~i+j}=(i-j)$.\,After repeating the same algebra for $\bar{\Gamma}$\,-\,sector,\,one can say that the asymptotic symmetry algebra for the most general  boundary conditions of $\mathcal{N}=(1,1)$ supergravity  is two copies of the affine $\mathfrak{osp}(1|2)_k$ algebra as in Ref.\cite{Valcarcel:2018kwd}.
\subsection{$\mathfrak{osp}(1|2) \oplus \mathfrak{osp}(1|2)$  Chern\,-\,Simons $\mathcal{N}=(1,1)$ Supergravity for Superconformal boundary}\label{bhreduction1}
Under the following restrictions as the Drinfeld\,-\,Sokolov highest weight gauge condition,
\begin{eqnarray}
\mathcal{L}^0=\mathcal{G}^{+\frac{1}{2}}=0,\,\mathcal{L}^{-1}=\mathcal{L},\,\mathcal{G}^{-\frac{1}{2}}=\mathcal{G},\,\alpha_{+1}\mathcal{L}^{+1}=1
\end{eqnarray}
on the boundary conditions with the $\mathfrak{osp}(1|2)$ superalgebra valued connection (\ref{bouncond999}),\,one can get the superconformal boundary conditions as the supersymmetric extension of the Brown\,-\,Henneaux boundary conditions proposed in \cite{Brown:1986nw} for $AdS_3$ supergravity.\,Therefore we have the supersymmetric connection as,
\begin{eqnarray}\label{bouncond9999}
a_\varphi &=&
 \Lt_{1}
+\alpha_{-1}\mathcal{L} \Lt_{-1}
+\beta_{-\frac{1}{2}}\mathcal{G}\Gt_{-\frac{1}{2}},\\
a_t &=&
\mathcal{\mu} \Lt_{1}
+\sum_{i=-1}^{0}\mathcal{\mu}^i \Lt_{i}
+\mathcal{\nu}\Gt_{{\frac{1}{2}}}
+\mathcal{\nu}^{-{\frac{1}{2}}}\Gt_{-{\frac{1}{2}}}
\end{eqnarray}
where $\mathcal{\mu}\equiv\mathcal{\mu}^{+1}$\,and\,$\mathcal{\nu}\equiv\mathcal{\nu}^{+{\frac{1}{2}}}$ can be interpreted as an independent $chemical$ $potentials$.\,This means that we assume the chemical potential to be fixed at infinity,\,i.e.\,$\delta\mu=0$.\,The functions $\mu^{0}$,\,$\mu^{-1}$ and $\mathcal{\nu}^{-{\frac{1}{2}}}$  are fixed by the flatness condition (\ref{flatness}) as
\begin{eqnarray}
\mu^{0}&=&-\mu',\\
\mu^{-1}&=&\frac{1}{2}\mu''+\alpha_{-1}\mathcal{L} \mu+\beta_{-\frac{1}{2}}\mathcal{G}\nu,\\
\mathcal{\nu}^{-{\frac{1}{2}}}&=&-\nu' +\beta_{-\frac{1}{2}}\mathcal{G}\nu.
\end{eqnarray}
For the fixed chemical potentials $\mu$ and $\nu$,\,the time evolution of canonical boundary charges $\cL$ and $\cG$ can be written as
\begin{eqnarray}
\partial_{t}\mathcal{L}	&=&-\frac{\mu'''}{2\alpha_{-1}} +2\mathcal{L}\mu'+\mathcal{L}'\mu+\frac{\beta_{-{\frac{1}{2}}}}{\alpha_{-1}}\bigg(3\mathcal{G}\nu'+\mathcal{G}'\nu\bigg),\label{bc14a}\\
\partial_{t}\mathcal{G} &=&-\frac{\nu''}{\alpha_{-1}} -\frac{\alpha_{-1}}{\beta_{-{\frac{1}{2}}}} \mathcal{L}\nu+\frac{3}{2}\mu'\mathcal{G}+\mu \mathcal{G}'\label{bouncon222}
\end{eqnarray}
where $\alpha_{-1}$ and $\beta_{-{\frac{1}{2}}}$ are some scaling parameters to be determined later.\,Now,\,we are in a position to work  the superconformal asymptotic symmetry algebra under the Drinfeld\,-\,Sokolov reduction.\,This reduction implies that the only independent parameters $\mathcal{\epsilon}\equiv\mathcal{\epsilon}^{+1}$,\,and $\mathcal{\zeta}\equiv\mathcal{\zeta}^{+{\frac{1}{2}}}$.\,One can start to compute the gauge transformations for asymptotic symmetry algebra by considering all transformations $(\ref{boundarycond})$ that preserve the boundary conditions.\,with the gauge parameter in the $\mathfrak{osp}(1|2)$ superalgebra $(\ref{boundarycond2})$ as
\begin{equation}
\lambda\,=\,b^{-1}\bigg[\mathcal{\epsilon} \Lt_{1}-\epsilon' \Lt_{0}
+\bigg(\frac{1}{2}\epsilon''+\alpha_{-1}\mathcal{L} \epsilon+\beta_{-\frac{1}{2}}\mathcal{G}\zeta\bigg)\Lt_{-1}
+\zeta\Gt_{+\frac{1}{2}}
+\bigg(\beta_{-\frac{1}{2}}\mathcal{G}\epsilon-\zeta'\bigg)\Gt_{-\frac{1}{2}}
 \bigg]b.
\end{equation}
The flat connection condition $(\ref{boundarycond})$ with $\alpha_{-1}=\frac{6}{c}$ and $\beta_{-{\frac{1}{2}}}=-\frac{3}{c}$ becomes:
\begin{eqnarray}
\delta_{\lambda}\mathcal{L}	&=&\frac{c}{12}\epsilon'''+2\mathcal{L}\epsilon'+\mathcal{L}'\epsilon-\frac{1}{2}\bigg(3\mathcal{G}\zeta'+\mathcal{G}'\zeta\bigg),\label{bc33}\\
\delta_{\lambda}\mathcal{G} &=&\frac{c}{3}\zeta''+2\mathcal{L}\zeta+\frac{3}{2}\epsilon'\mathcal{G}+\epsilon \mathcal{G}'.\label{bouncon22244}
\end{eqnarray}
The variation of  canonical boundary charge $\mathcal{Q[\lambda]}$ $(\ref{Qvar})$  can be integrated which reads
\begin{equation}
\mathcal{Q[\lambda]}=\int\mathrm{d}\varphi\;\left[\mathcal{L}\epsilon+\mathcal{G}\zeta\right].\label{boundaryco11111}
\end{equation}
This leads to operator product expansions in the complex coordinates by using $(\ref{Qvar2})$
\begin{eqnarray}\label{ope}
&\mathcal{L}(z_1)\mathcal{L}(z_2)\,\sim  \,{{c\over 2}\over{z_{12}^{4}}}\,+\, {2\,\mathcal{L}\over{z_{12}^{2}}}\, + \,{ \mathcal{L}'\over{z_{12}}},\\
&\mathcal{L}(z_1)\mathcal{G}(z_2)\, \sim  \,{\frac{3}{2}\mathcal{G}\over{z_{12}^{2}}}\, + \,{ \mathcal{G}'\over{z_{12}}},~~~
\mathcal{G}(z_1)\mathcal{G}(z_2)\,\sim \,{{2c\over 3}\over{z_{12}^{3}}}\,+\, {2\,\mathcal{L}\over{z_{12}}}.
\end{eqnarray}
After repeating the same algebra for $\bar{\Gamma}$\,-\,sector,\,one can say that the asymptotic symmetry algebra for a set of boundary conditions of $\mathcal{N}=(1,1)$ supergravity  is two copies of the super\,-\,Virasoro algebra with central charce $c\,=\,6k$.


\section{$\mathcal{N}=(1,1)$ $\mathfrak{osp}(3|2) \oplus \mathfrak{osp}(3|2)$ Higher\,-\,spin  Chern\,-\,Simons Supergravity}

\subsection{For Affine Boundary}
In this section,\,we will construct an extension of the $\mathcal{N}=(1,1)$ higher\,-\,spin supergravity theory.\,To this end,\,$\mathfrak{osp}(3|2)$ algebra is defined by using  an higher\,-\,spin extension of $\mathfrak{osp}(1|2)$ algebra which can be presented as a sub\,-\,superalgebra as the ordinary case of $\mathfrak{osp}(N|M)$ gauge algebra.
If $\Lt_{i}\,(i=\pm1,0)$,\,$\Gt_{p}\,(p=\pm\frac{1}{2})$,\,$\At_{i}\,(i=\pm1,0)$,\,and $\St_{p}\,(p=\pm\frac{1}{2},\pm\frac{3}{2})$are the generators of $\mathfrak{osp}(3|2)$ superalgebra,\,we have expressed $\mathfrak{osp}(3|2)$ superalgebra such that
\begin{eqnarray}\label{algebraLGAS}
\left[\Lt_{i},\Lt_{j}\right]&=& \left(i-j\right)\Lt_{i+j},\,~~,\label{algebraLGASs}
\left[\Lt_{i},\Gt_{p}\right]=\big(\frac{i}{2}-p\big)\Gt_{i+p},\,\\
\left\{\Gt_{p},\Gt_{q}\right\}&=&\sigma_{1}\At_{p+q}+\sigma_{2}\Lt_{p+q},\,~~
\left\{\Gt_{p},\St_{q}\right\}=\bigg(\frac{3p}{2}-\frac{q}{2}\bigg)\bigg(\sigma_{3}\At_{p+q}+\sigma_{4}\Lt_{p+q}\bigg),\,\\
\left[\Lt_{i},\At_{j}\right]&=& \left(\frac{i}{2}-j\right)\At_{i+j},\,~~
\left[\Lt_{i},\St_{p}\right]=\bigg(\frac{3i}{2}-p\bigg)\St_{i+p},\,\\
\left[\At_{i},\At_{j}\right]&=&\left(i-j\right)\left(\sigma_{5}\Lt_{i+j}+\sigma_{6}\At_{i+j}\right),\,~~
\left[\At_{i},\Gt_{p}\right]=\sigma_{7}\St_{i+p}+\sigma_{8}\bigg(\frac{i}{2}-p\bigg)\Gt_{i+p},\\
\left[\At_{i},\St_{p}\right]&=&\sigma_{9}\bigg(\frac{3i}{2}-p\bigg)\St_{i+p}+\sigma_{10}\bigg(3 i^2-2ip+p^2-\frac{9}{4}\bigg)\Gt_{i+p},\,~~\\
\left\{\St_{p},\St_{q}\right\}&=&\bigg(3p^2-4pq+3q^2-\frac{9}{2}\bigg)\bigg(\sigma_{11}\At_{p+q}+\sigma_{12}\Lt_{p+q}\bigg).\label{algebraLGASss}
\end{eqnarray}
The super Jacobi identities  give us the nontrivial relations for some constants $\sigma_{i}'s,(i=1,2,\dots,12)$ appearing on the RHS of eq.$(\ref{algebraLGASs})-(\ref{algebraLGASss})$ as,
\begin{eqnarray}
\sigma _1\,=\, 0,\,~~
\sigma _2\,&=&\, -\sigma _3 \sigma _7,\,~~
\sigma _4\,=\, -\sigma _3 \sigma _8,\,~~
\sigma _5\,=\, \frac{1}{2} \left(\sigma _8-2 \sigma _9\right) \left(\sigma _8+\sigma _9\right),\,~~
\sigma _6\,=\, \frac{1}{2} \left(5 \sigma _9-\sigma _8\right),\nn\\
\sigma _{10}\,&=&\, -\frac{\left(\sigma _8-\sigma _9\right){}^2}{4 \sigma _7},\,~~
\sigma _{11}\,=\, -\frac{\sigma_3 \left(\sigma _8-\sigma _9\right)}{4 \sigma _7},\,~~
\sigma _{12}\,=\,-\frac{\sigma _3 \sigma _9 \left(\sigma _9-\sigma _8\right)}{4 \sigma _7}.
\end{eqnarray}
For the corresponding algebra, the resulting relations are as $
\sigma _3\,=\, 2,\,~~\sigma _7\,=\, -1,\,~~\sigma _8\,=\, 0,\,~~\sigma _9\,=\,1$.\,We are now ready to formulate most general boundary conditions for asymptotically $AdS_3$ spacetimes:
\begin{eqnarray}\label{bouncondosp32}
a_\varphi &=&
 \alpha_i\mathcal{L}^i \Lt_{i}
+\gamma_i\mathcal{A}^i \At_{i}
+\beta_p\mathcal{G}^{p}\Gt_{p}
+\tau_p\mathcal{S}^{p}\St_{p}
,\\
a_t &=&
 \mu^i \Lt_{i}
+\chi^i \At_{i}
+\mathit{f}^{p}\Gt_{p}
+\nu^{p}\St_{p}
\end{eqnarray}
where $\left( \alpha_i,\gamma_i,\beta_p,\tau_p\right)$'s are some scaling parameters to be determined later and we have twelve functions: six bosonic $\left(\mathcal{L}^i,\mathcal{A}^i\right)$ and six fermionic $\left(\mathcal{G}^{p},\mathcal{S}^{p}\right)$ as the $charges$ and also here, the time component has in total of twelve independent functions $( \mu^i,\chi^i,\mathit{f}^{p},\nu^{p})$
 as the $chemical~~potentials$.\\ The flat connection conditions $(\ref{flatness})$ for fixed chemical potentials impose the following additional conditions as the temporal evolution of the twelve independent source fields,\,$(\mathcal{L}^i,\mathcal{A}^i,\mathcal{G}^{p},\mathcal{S}^{p})$ as
\begin{eqnarray}
\alpha_{\pm1}\partial_{t}\mathcal{L}^{\pm1}&=&
\partial_{\varphi}\mu^{\pm1}
\pm \alpha_{\pm1}\mu^0\mathcal{L}^{\pm1}
\mp \alpha_0\mu^{\pm1}\mathcal{L}^0
+2\beta_{\pm\frac{1}{2}}\mathcal{G}^{\pm\frac{1}{2}}\nu^{\pm\frac{1}{2}}
\mp\mathcal{A}^{\pm1}\gamma_{\pm1}\chi^0
\pm\mathcal{A}^0\gamma_0\chi^{\pm1}\nn\\
&\space&+3\tau_{\mp\frac{1}{2}}\mathit{f}^{\pm\frac{3}{2}}\mathcal{S}^{\mp\frac{1}{2}}
-2\tau_{\pm\frac{1}{2}}\mathit{f}^{\pm\frac{1}{2}}\mathcal{S}^{\pm\frac{1}{2}}
+3\tau_{\pm\frac{3}{2}}\mathit{f}^{\mp\frac{1}{2}}\mathcal{S}^{\pm\frac{3}{2}},\\
\frac{1}{2}\alpha_0\partial_{t}\mathcal{L}^0&=&
\frac{1}{2}\partial_{\varphi}\mu^0
+\alpha_{+1}\mu^{-1}\mathcal{L}^1
-\alpha_{-1}\mu^{+1}\mathcal{L}^{-1}
+\beta_{+\frac{1}{2}}\mathcal{G}^{+\frac{1}{2}}\nu^{-\frac{1}{2}}
+\beta_{-\frac{1}{2}}\mathcal{G}^{-\frac{1}{2}}\nu^{+\frac{1}{2}}\nn\\
&\space&-\mathcal{A}^{+1}\gamma_{+1}\chi^{-1}
+\mathcal{A}^{-1}\gamma_{-1}\chi^{+1}
+\frac{9}{2}\tau_{-\frac{3}{2}}\mathit{f}^{+\frac{3}{2}}\mathcal{S}^{-\frac{3}{2}}
-\frac{1}{2}\tau_{-\frac{1}{2}}\mathit{f}^{+\frac{1}{2}}\mathcal{S}^{-\frac{1}{2}}\nn\\
&\space&-\frac{1}{2}\tau_{+\frac{1}{2}}\mathit{f}^{-\frac{1}{2}}\mathcal{S}^{+\frac{1}{2}}
+\frac{9}{2}\tau_{+\frac{3}{2}}\mathit{f}^{-\frac{3}{2}}\mathcal{S}^{+\frac{3}{2}},\\
\mathit{g}_{\pm\frac{1}{2}}\partial_{t}\mathcal{G}^{\pm\frac{1}{2}}{}&=&
\partial_{\varphi}\nu^{\pm\frac{1}{2}}
\mp\frac{1}{2}\alpha_0\nu^{\pm\frac{1}{2}}\mathcal{L}^0
\pm\alpha_{\pm1}\nu^{\mp\frac{1}{2}}\mathcal{L}^{\pm1}
\mp\beta_{\mp\frac{1}{2}}\mathcal{G}^{\mp\frac{1}{2}}\mu^{\pm1}
\pm\frac{1}{2}\beta_{\pm\frac{1}{2}}\mathcal{G}^{\pm\frac{1}{2}}\mu^0\nn\\
&\space&+\frac{3}{2}\mathcal{A}^{\mp1}\gamma_{\mp1}\mathit{f}^{\pm\frac{3}{2}}
-\frac{1}{2}\mathcal{A}^0\gamma_0\mathit{f}^{\pm\frac{1}{2}}
+\frac{1}{2}\mathcal{A}^{\pm1}\gamma_{\pm1}\mathit{f}^{\mp\frac{1}{2}}\nn\\
&\space&-\frac{3}{2}\tau_{\pm\frac{3}{2}}\mathcal{S}^{\pm\frac{3}{2}}\chi^{\mp1}
+\frac{1}{2}\tau_{\pm\frac{1}{2}}\mathcal{S}^{\pm\frac{1}{2}}\chi^0
-\frac{1}{2}\tau_{\mp\frac{1}{2}}\mathcal{S}^{\mp\frac{1}{2}}\chi^{\pm1},\\
\frac{1}{2} \gamma _0 \partial_{t}\mathcal{A}^0&=&
\frac{1}{2}\partial_{\varphi}\chi^0+\mathcal{A}^{+1}\gamma_{+1}\mu^{-1}-\mathcal{A}^{-1}\gamma_{-1}\mu^{+1}+\frac{5}{2}\mathcal{A}^{+1}\gamma_{+1}\chi^{-1}-\frac{5}{2}\mathcal{A}^{-1}
\gamma_{-1}\chi^{+1}-\beta_{-\frac{1}{2}}\mathit{f}^{+\frac{1}{2}}\mathcal{G}^{-\frac{1}{2}}\nn\\
&\space&+\beta_{+\frac{1}{2}}\mathit{f}^{-\frac{1}{2}}
\mathcal{G}^{+\frac{1}{2}}-\frac{9}{2}\tau_{-\frac{3}{2}}\mathit{f}^{+\frac{3}{2}}\mathcal{S}^{-\frac{3}{2}}+\frac{1}{2}\tau_{-\frac{1}{2}}\mathit{f}^{+\frac{1}{2}}
\mathcal{S}^{-\frac{1}{2}}+\frac{1}{2}\tau_{+\frac{1}{2}}\mathit{f}^{-\frac{1}{2}}\mathcal{S}^{+\frac{1}{2}}-\frac{9}{2}\tau_{+\frac{3}{2}}\mathit{f}^{-\frac{3}{2}}
\mathcal{S}^{+\frac{3}{2}}\nn\\
&\space&-\tau_{+\frac{1}{2}}\nu^{-\frac{1}{2}}\mathcal{S}^{+\frac{1}{2}}+\tau_{-\frac{1}{2}}\nu^{+\frac{1}{2}}\mathcal{S}^{-\frac{1}{2}}+\alpha_{+1}
\chi^{-1}\mathcal{L}^{+1}-\alpha_{-1}\chi^{+1}\mathcal{L}^{-1},\\
\gamma _{\pm1} \partial_{t}\mathcal{A}^{\pm1}&=&
\partial_{\varphi}\chi^{+1}
\pm\mathcal{A}^{\pm1}\gamma_{\pm1}\mu^0
\mp\mathcal{A}^0\gamma_0\mu^{\pm1}
\pm\frac{5}{2}\mathcal{A}^{\pm1}\gamma_{\pm1}\chi^0
\mp\frac{5}{2}\mathcal{A}^0\gamma_0\chi^{\pm1}
\mp3\beta_{\mp\frac{1}{2}}\mathit{f}^{\pm\frac{3}{2}}\mathcal{G}^{-\frac{1}{2}}\nn\\
&\space&+\beta_{\pm\frac{1}{2}}\mathit{f}^{\pm\frac{1}{2}}\mathcal{G}^{\pm\frac{1}{2}}
-3\tau_{\mp\frac{1}{2}}\mathit{f}^{\pm\frac{3}{2}}\mathcal{S}^{\mp\frac{1}{2}}
+2\tau_{\pm\frac{1}{2}}\mathit{f}^{\pm\frac{1}{2}}\mathcal{S}^{\pm\frac{1}{2}}
-3\tau_{\pm\frac{3}{2}}\mathit{f}^{\mp\frac{1}{2}}\mathcal{S}^{\pm\frac{3}{2}}
-3\tau_{\pm\frac{3}{2}}\nu^{\mp\frac{1}{2}}\mathcal{S}^{\pm\frac{3}{2}}\nn\\
&\space&+\tau_{\pm\frac{1}{2}}\nu^{\pm\frac{1}{2}}\mathcal{S}^{\pm\frac{1}{2}}
\pm \alpha_{\pm1}\chi^0\mathcal{L}^{\pm1}
\mp\alpha_0\chi^{\pm1}\mathcal{L}^0,\\
\tau_{\pm\frac{1}{2}} \partial_{t}\mathcal{S}^{\pm\frac{1}{2}}&=&
\partial_{\varphi}\mathit{f}^{\pm\frac{1}{2}}
-\mathcal{A}^{\pm1}\gamma_{\pm1}\nu^{\mp\frac{1}{2}}
-\mathcal{A}^0\gamma_0\nu^{\pm\frac{1}{2}}
\mp3\mathcal{A}^{\mp1}\gamma_{\mp1}\mathit{f}^{\pm\frac{3}{2}}
\mp\frac{1}{2}\mathcal{A}^0\gamma_0\mathit{f}^{\pm\frac{1}{2}}
\pm2\mathcal{A}^{\pm1}\gamma_{\pm1}\mathit{f}^{\mp\frac{1}{2}}\nn\\
&\space&\mp3\alpha_{\mp1}\mathit{f}^{\pm\frac{3}{2}}\mathcal{L}^{\mp1}
-\frac{1}{2}\alpha_0\mathit{f}^{\pm\frac{1}{2}}\mathcal{L}^0
\pm2\alpha_{\pm1}\mathit{f}^{\mp\frac{1}{2}}\mathcal{L}^{\pm1}
+\beta_{\pm\frac{1}{2}}\mathcal{G}^{\pm\frac{1}{2}}\chi^0
+\beta_{\mp\frac{1}{2}}\mathcal{G}^{\mp\frac{1}{2}}\chi^{\pm1}\nn\\
&\space&\pm3\tau_{\pm\frac{3}{2}}\mu^{\mp1}\mathcal{S}^{\pm\frac{3}{2}}
\pm\frac{1}{2}\tau_{\pm\frac{1}{2}}\mu^0\mathcal{S}^{\pm\frac{1}{2}}
\mp2\tau_{\mp\frac{1}{2}}\mu^{\pm1}\mathcal{S}^{\mp\frac{1}{2}}
\pm\tau_{\pm\frac{3}{2}}\mathcal{S}^{\pm\frac{3}{2}}\chi^{\mp1}
\pm\frac{1}{2}\tau_{\pm\frac{1}{2}}\mathcal{S}^{\pm\frac{1}{2}}\chi^0\nn\\
&\space&\mp2\tau_{\mp\frac{1}{2}}\mathcal{S}^{\mp\frac{1}{2}}\chi^{\pm1},\\
\tau_{\pm\frac{3}{2}} \partial_{t}\mathcal{S}^{\pm\frac{3}{2}}&=&
 \partial_{\varphi}\mathit{f}^{\pm\frac{3}{2}}
-\mathcal{A}^{\pm1}\gamma_{\pm1}\nu^{\pm\frac{1}{2}}
\mp\frac{3}{2}\mathcal{A}^0\gamma_0\mathit{f}^{\pm\frac{3}{2}}
\pm\mathcal{A}^{\pm1}\gamma_{\pm1}\mathit{f}^{\pm\frac{1}{2}}
\mp\frac{3}{2}\alpha_0\mathit{f}^{\pm\frac{3}{2}}\mathcal{L}^0
+\alpha_{\pm1}\mathit{f}^{\pm\frac{1}{2}}\mathcal{L}^{\pm1}\nn\\
&\space&+\beta_{\pm\frac{1}{2}}\mathcal{G}^{\pm\frac{1}{2}}\chi^{\pm1}
\pm\frac{3}{2}\tau_{\pm\frac{3}{2}}\mu^0\mathcal{S}^{\pm\frac{3}{2}}
-\tau_{\pm\frac{1}{2}}\mu^{\pm1}\mathcal{S}^{\pm\frac{1}{2}}
\pm\frac{3}{2}\tau_{\pm\frac{3}{2}}\mathcal{S}^{\pm\frac{3}{2}}\chi^0
\mp\tau_{\pm\frac{1}{2}}\mathcal{S}^{\pm\frac{1}{2}}\chi^{\pm1}.
\end{eqnarray}\newpage
After the temporal evolution of the source fields,\,one can start to compute the gauge transformations for asymptotic symmetry algebra
by considering all transformations $(\ref{boundarycond})$ that preserve the boundary conditions,\,with the gauge parameter $\lambda$ in the $\mathfrak{osp}(3|2)$ superalgebra as,
\begin{equation}
    \lambda=b^{-1}
    \bigg[
 {\epsilon}^i \Lt_{i}
+{\kappa}^i \At_{i}
+{\zeta}^{p}\Gt_{p}
+{\varrho}^{p}\St_{p}
   \bigg]b.\label{boundarycond2221}
\end{equation}
Here,\,the gauge parameter has in total of twelve arbitrary  functions\,:six bosonic $(\mathcal{\epsilon}^i,\,{\kappa}^i)$ and six fermionic $(\mathcal{\zeta}^{p},\,\mathcal{\varrho}^{p})$ on the boundary.\,The condition $(\ref{boundarycond})$ impose that transformations on the gauge are given by
\begin{eqnarray}
\alpha_{\pm1}\delta_{\lambda}\mathcal{L}^{\pm1}&=&
 \partial_{\varphi}\epsilon^{\pm1}
\pm \alpha_{\pm1}\epsilon^0\mathcal{L}^{\pm1}
\mp \alpha_0\epsilon^{\pm1}\mathcal{L}^0
+2\beta_{\pm\frac{1}{2}}\mathcal{G}^{\pm\frac{1}{2}}\varrho^{\pm\frac{1}{2}}
\mp\mathcal{A}^{\pm1}\gamma_{\pm1}\kappa^0
\pm\mathcal{A}^0\gamma_0\kappa^{\pm1}\nn\\
&\space&+3\tau_{\mp\frac{1}{2}}\zeta^{\pm\frac{3}{2}}\mathcal{S}^{\mp\frac{1}{2}}
-2\tau_{\pm\frac{1}{2}}\zeta^{\pm\frac{1}{2}}\mathcal{S}^{\pm\frac{1}{2}}
+3\tau_{\pm\frac{3}{2}}\zeta^{\mp\frac{1}{2}}\mathcal{S}^{\pm\frac{3}{2}},\label{tra211}\\
\frac{1}{2}\alpha_0\delta_{\lambda}\mathcal{L}^0&=&
\frac{1}{2}\partial_{\varphi}\epsilon^0
+\alpha_{+1}\epsilon^{-1}\mathcal{L}^1
-\alpha_{-1}\epsilon^{+1}\mathcal{L}^{-1}
+\beta_{+\frac{1}{2}}\mathcal{G}^{+\frac{1}{2}}\varrho^{-\frac{1}{2}}
+\beta_{-\frac{1}{2}}\mathcal{G}^{-\frac{1}{2}}\varrho^{+\frac{1}{2}}\nn\\
&\space&-\mathcal{A}^{+1}\gamma_{+1}\kappa^{-1}
+\mathcal{A}^{-1}\gamma_{-1}\kappa^{+1}
+\frac{9}{2}\tau_{-\frac{3}{2}}\zeta^{+\frac{3}{2}}\mathcal{S}^{-\frac{3}{2}}
-\frac{1}{2}\tau_{-\frac{1}{2}}\zeta^{+\frac{1}{2}}\mathcal{S}^{-\frac{1}{2}}\nn\\
&\space&-\frac{1}{2}\tau_{+\frac{1}{2}}\zeta^{-\frac{1}{2}}\mathcal{S}^{+\frac{1}{2}}
+\frac{9}{2}\tau_{+\frac{3}{2}}\zeta^{-\frac{3}{2}}\mathcal{S}^{+\frac{3}{2}},\label{tra212}\\
\mathit{g}_{\pm\frac{1}{2}}\delta_{\lambda}\mathcal{G}^{\pm\frac{1}{2}}{}&=&
\partial_{\varphi}\varrho^{\pm\frac{1}{2}}
\mp\frac{1}{2}\alpha_0\varrho^{\pm\frac{1}{2}}\mathcal{L}^0
\pm\alpha_{\pm1}\varrho^{\mp\frac{1}{2}}\mathcal{L}^{\pm1}
\mp\beta_{\mp\frac{1}{2}}\mathcal{G}^{\mp\frac{1}{2}}\epsilon^{\pm1}
\pm\frac{1}{2}\beta_{\pm\frac{1}{2}}\mathcal{G}^{\pm\frac{1}{2}}\epsilon^0\nn\\
&\space&+\frac{3}{2}\mathcal{A}^{\mp1}\gamma_{\mp1}\zeta^{\pm\frac{3}{2}}
-\frac{1}{2}\mathcal{A}^0\gamma_0\zeta^{\pm\frac{1}{2}}
+\frac{1}{2}\mathcal{A}^{\pm1}\gamma_{\pm1}\zeta^{\mp\frac{1}{2}}\nn\\
&\space&-\frac{3}{2}\tau_{\pm\frac{3}{2}}\mathcal{S}^{\pm\frac{3}{2}}\kappa^{\mp1}
+\frac{1}{2}\tau_{\pm\frac{1}{2}}\mathcal{S}^{\pm\frac{1}{2}}\kappa^0
-\frac{1}{2}\tau_{\mp\frac{1}{2}}\mathcal{S}^{\mp\frac{1}{2}}\kappa^{\pm1},\label{tra213}\\
\frac{1}{2} \gamma _0 \delta_{\lambda}\mathcal{A}^0&=&
\frac{1}{2}\partial_{\varphi}\kappa^0+\mathcal{A}^{+1}\gamma_{+1}\epsilon^{-1}-\mathcal{A}^{-1}\gamma_{-1}\epsilon^{+1}+\frac{5}{2}\mathcal{A}^{+1}\gamma_{+1}\kappa^{-1}-\frac{5}{2}\mathcal{A}^{-1}
\gamma_{-1}\kappa^{+1}-\beta_{-\frac{1}{2}}\zeta^{+\frac{1}{2}}\mathcal{G}^{-\frac{1}{2}}\nn\\
&\space&+\beta_{+\frac{1}{2}}\zeta^{-\frac{1}{2}}
\mathcal{G}^{+\frac{1}{2}}-\frac{9}{2}\tau_{-\frac{3}{2}}\zeta^{+\frac{3}{2}}\mathcal{S}^{-\frac{3}{2}}+\frac{1}{2}\tau_{-\frac{1}{2}}\zeta^{+\frac{1}{2}}
\mathcal{S}^{-\frac{1}{2}}+\frac{1}{2}\tau_{+\frac{1}{2}}\zeta^{-\frac{1}{2}}\mathcal{S}^{+\frac{1}{2}}-\frac{9}{2}\tau_{+\frac{3}{2}}\zeta^{-\frac{3}{2}}
\mathcal{S}^{+\frac{3}{2}}\nn\\
&\space&-\tau_{+\frac{1}{2}}\varrho^{-\frac{1}{2}}\mathcal{S}^{+\frac{1}{2}}+\tau_{-\frac{1}{2}}\varrho^{+\frac{1}{2}}\mathcal{S}^{-\frac{1}{2}}+\alpha_{+1}
\kappa^{-1}\mathcal{L}^{+1}-\alpha_{-1}\kappa^{+1}\mathcal{L}^{-1},\label{tra214}\\
\gamma _{\pm1} \delta_{\lambda}\mathcal{A}^{\pm1}&=&
\partial_{\varphi}\kappa^{+1}
\pm\mathcal{A}^{\pm1}\gamma_{\pm1}\epsilon^0
\mp\mathcal{A}^0\gamma_0\epsilon^{\pm1}
\pm\frac{5}{2}\mathcal{A}^{\pm1}\gamma_{\pm1}\kappa^0
\mp\frac{5}{2}\mathcal{A}^0\gamma_0\kappa^{\pm1}
\mp3\beta_{\mp\frac{1}{2}}\zeta^{\pm\frac{3}{2}}\mathcal{G}^{-\frac{1}{2}}\nn\\
&\space&\pm\beta_{\pm\frac{1}{2}}\zeta^{\pm\frac{1}{2}}\mathcal{G}^{\pm\frac{1}{2}}
-3\tau_{\mp\frac{1}{2}}\zeta^{\pm\frac{3}{2}}\mathcal{S}^{\mp\frac{1}{2}}
+2\tau_{\pm\frac{1}{2}}\zeta^{\pm\frac{1}{2}}\mathcal{S}^{\pm\frac{1}{2}}
-3\tau_{\pm\frac{3}{2}}\zeta^{\mp\frac{1}{2}}\mathcal{S}^{\pm\frac{3}{2}}
-3\tau_{\pm\frac{3}{2}}\varrho^{\mp\frac{1}{2}}\mathcal{S}^{\pm\frac{3}{2}}\nn\\
&\space&+\tau_{\pm\frac{1}{2}}\varrho^{\pm\frac{1}{2}}\mathcal{S}^{\pm\frac{1}{2}}
\pm \alpha_{\pm1}\kappa^0\mathcal{L}^{\pm1}
\mp\alpha_0\kappa^{\pm1}\mathcal{L}^0,\label{tra215}\\
\tau_{\pm\frac{1}{2}} \delta_{\lambda}\mathcal{S}^{\pm\frac{1}{2}}&=&
\partial_{\varphi}\zeta^{\pm\frac{1}{2}}
-\mathcal{A}^{\pm1}\gamma_{\pm1}\varrho^{\mp\frac{1}{2}}
-\mathcal{A}^0\gamma_0\varrho^{\pm\frac{1}{2}}
\mp3\mathcal{A}^{\mp1}\gamma_{\mp1}\zeta^{\pm\frac{3}{2}}
\mp\frac{1}{2}\mathcal{A}^0\gamma_0\zeta^{\pm\frac{1}{2}}
\pm2\mathcal{A}^{\pm1}\gamma_{\pm1}\zeta^{\mp\frac{1}{2}}\nn\\
&\space&\mp3\alpha_{\mp1}\zeta^{\pm\frac{3}{2}}\mathcal{L}^{\mp1}
-\frac{1}{2}\alpha_0\zeta^{\pm\frac{1}{2}}\mathcal{L}^0
\pm2\alpha_{\pm1}\zeta^{\mp\frac{1}{2}}\mathcal{L}^{\pm1}
+\beta_{\pm\frac{1}{2}}\mathcal{G}^{\pm\frac{1}{2}}\kappa^0
+\beta_{\mp\frac{1}{2}}\mathcal{G}^{\mp\frac{1}{2}}\kappa^{\pm1}\nn\\
&\space&\pm3\tau_{\pm\frac{3}{2}}\epsilon^{\mp1}\mathcal{S}^{\pm\frac{3}{2}}
\pm\frac{1}{2}\tau_{\pm\frac{1}{2}}\epsilon^0\mathcal{S}^{\pm\frac{1}{2}}
\mp2\tau_{\mp\frac{1}{2}}\epsilon^{\pm1}\mathcal{S}^{\mp\frac{1}{2}}
\pm\tau_{\pm\frac{3}{2}}\mathcal{S}^{\pm\frac{3}{2}}\kappa^{\mp1}
\pm\frac{1}{2}\tau_{\pm\frac{1}{2}}\mathcal{S}^{\pm\frac{1}{2}}\kappa^0\nn\\
&\space&\mp2\tau_{\mp\frac{1}{2}}\mathcal{S}^{\mp\frac{1}{2}}\kappa^{\pm1},\label{tra216}
\end{eqnarray}
\begin{eqnarray}
\tau_{\pm\frac{3}{2}} \delta_{\lambda}\mathcal{S}^{\pm\frac{3}{2}}&=&
 \partial_{\varphi}\zeta^{\pm\frac{3}{2}}
-\mathcal{A}^{\pm1}\gamma_{\pm1}\varrho^{\pm\frac{1}{2}}
\mp\frac{3}{2}\mathcal{A}^0\gamma_0\zeta^{\pm\frac{3}{2}}
\pm\mathcal{A}^{\pm1}\gamma_{\pm1}\zeta^{\pm\frac{1}{2}}
\mp\frac{3}{2}\alpha_0\zeta^{\pm\frac{3}{2}}\mathcal{L}^0
+\alpha_{\pm1}\zeta^{\pm\frac{1}{2}}\mathcal{L}^{\pm1}\nn\\
&\space&+\beta_{\pm\frac{1}{2}}\mathcal{G}^{\pm\frac{1}{2}}\kappa^{\pm1}
\pm\frac{3}{2}\tau_{\pm\frac{3}{2}}\epsilon^0\mathcal{S}^{\pm\frac{3}{2}}
-\tau_{\pm\frac{1}{2}}\epsilon^{\pm1}\mathcal{S}^{\pm\frac{1}{2}}\pm\frac{3}{2}\tau_{\pm\frac{3}{2}}\mathcal{S}^{\pm\frac{3}{2}}\kappa^0\nn\\
&\space&\mp\tau_{\pm\frac{1}{2}}\mathcal{S}^{\pm\frac{1}{2}}\kappa^{\pm1}.\label{tra217}
\end{eqnarray}
With analogical reasoning,\,they obey the following transformations
\begin{eqnarray}
\alpha_{\pm1}\delta_{\lambda}\mu^{\pm1}&=&
 \partial_{t}\epsilon^{\pm1}
\pm \alpha_{\pm1}\epsilon^0\mu^{\pm1}
\mp \alpha_0\epsilon^{\pm1}\mu^0
+2\beta_{\pm\frac{1}{2}}\mathit{f}^{\pm\frac{1}{2}}\varrho^{\pm\frac{1}{2}}
\mp\chi^{\pm1}\gamma_{\pm1}\kappa^0
\pm\chi^0\gamma_0\kappa^{\pm1}\nn\\
&\space&+3\tau_{\mp\frac{1}{2}}\zeta^{\pm\frac{3}{2}}\nu^{\mp\frac{1}{2}}
-2\tau_{\pm\frac{1}{2}}\zeta^{\pm\frac{1}{2}}\nu^{\pm\frac{1}{2}}
+3\tau_{\pm\frac{3}{2}}\zeta^{\mp\frac{1}{2}}\nu^{\pm\frac{3}{2}},\\
\frac{1}{2}\alpha_0\delta_{\lambda}\mu^0&=&
\frac{1}{2}\partial_{t}\epsilon^0
+\alpha_{+1}\epsilon^{-1}\mu^1
-\alpha_{-1}\epsilon^{+1}\mu^{-1}
+\beta_{+\frac{1}{2}}\mathit{f}^{+\frac{1}{2}}\varrho^{-\frac{1}{2}}
+\beta_{-\frac{1}{2}}\mathit{f}^{-\frac{1}{2}}\varrho^{+\frac{1}{2}}\nn\\
&\space&-\chi^{+1}\gamma_{+1}\kappa^{-1}
+\chi^{-1}\gamma_{-1}\kappa^{+1}
+\frac{9}{2}\tau_{-\frac{3}{2}}\zeta^{+\frac{3}{2}}\nu^{-\frac{3}{2}}
-\frac{1}{2}\tau_{-\frac{1}{2}}\zeta^{+\frac{1}{2}}\nu^{-\frac{1}{2}}\nn\\
&\space&-\frac{1}{2}\tau_{+\frac{1}{2}}\zeta^{-\frac{1}{2}}\nu^{+\frac{1}{2}}
+\frac{9}{2}\tau_{+\frac{3}{2}}\zeta^{-\frac{3}{2}}\nu^{+\frac{3}{2}},\\
\mathit{g}_{\pm\frac{1}{2}}\delta_{\lambda}\mathit{f}^{\pm\frac{1}{2}}{}&=&
\partial_{t}\varrho^{\pm\frac{1}{2}}
\mp\frac{1}{2}\alpha_0\varrho^{\pm\frac{1}{2}}\mu^0
\pm\alpha_{\pm1}\varrho^{\mp\frac{1}{2}}\mu^{\pm1}
\mp\beta_{\mp\frac{1}{2}}\mathit{f}^{\mp\frac{1}{2}}\epsilon^{\pm1}
\pm\frac{1}{2}\beta_{\pm\frac{1}{2}}\mathit{f}^{\pm\frac{1}{2}}\epsilon^0\nn\\
&\space&+\frac{3}{2}\chi^{\mp1}\gamma_{\mp1}\zeta^{\pm\frac{3}{2}}
-\frac{1}{2}\chi^0\gamma_0\zeta^{\pm\frac{1}{2}}
+\frac{1}{2}\chi^{\pm1}\gamma_{\pm1}\zeta^{\mp\frac{1}{2}}\nn\\
&\space&-\frac{3}{2}\tau_{\pm\frac{3}{2}}\nu^{\pm\frac{3}{2}}\kappa^{\mp1}
+\frac{1}{2}\tau_{\pm\frac{1}{2}}\nu^{\pm\frac{1}{2}}\kappa^0
-\frac{1}{2}\tau_{\mp\frac{1}{2}}\nu^{\mp\frac{1}{2}}\kappa^{\pm1},\\
\frac{1}{2} \gamma _0 \delta_{\lambda}\chi^0&=&
\frac{1}{2}\partial_{t}\kappa^0+\chi^{+1}\gamma_{+1}\epsilon^{-1}-\chi^{-1}\gamma_{-1}\epsilon^{+1}+\frac{5}{2}\chi^{+1}\gamma_{+1}\kappa^{-1}-\frac{5}{2}\chi^{-1}
\gamma_{-1}\kappa^{+1}-\beta_{-\frac{1}{2}}\zeta^{+\frac{1}{2}}\mathit{f}^{-\frac{1}{2}}\nn\\
&\space&+\beta_{+\frac{1}{2}}\zeta^{-\frac{1}{2}}
\mathit{f}^{+\frac{1}{2}}-\frac{9}{2}\tau_{-\frac{3}{2}}\zeta^{+\frac{3}{2}}\nu^{-\frac{3}{2}}+\frac{1}{2}\tau_{-\frac{1}{2}}\zeta^{+\frac{1}{2}}
\nu^{-\frac{1}{2}}+\frac{1}{2}\tau_{+\frac{1}{2}}\zeta^{-\frac{1}{2}}\nu^{+\frac{1}{2}}-\frac{9}{2}\tau_{+\frac{3}{2}}\zeta^{-\frac{3}{2}}
\nu^{+\frac{3}{2}}\nn\\
&\space&-\tau_{+\frac{1}{2}}\varrho^{-\frac{1}{2}}\nu^{+\frac{1}{2}}+\tau_{-\frac{1}{2}}\varrho^{+\frac{1}{2}}\nu^{-\frac{1}{2}}+\alpha_{+1}
\kappa^{-1}\mu^{+1}-\alpha_{-1}\kappa^{+1}\mu^{-1},
\end{eqnarray}
\begin{eqnarray}
\gamma _{\pm1} \delta_{\lambda}\chi^{\pm1}&=&
\partial_{t}\kappa^{+1}
\pm\chi^{\pm1}\gamma_{\pm1}\epsilon^0
\mp\chi^0\gamma_0\epsilon^{\pm1}
\pm\frac{5}{2}\chi^{\pm1}\gamma_{\pm1}\kappa^0
\mp\frac{5}{2}\chi^0\gamma_0\kappa^{\pm1}
\mp3\beta_{\mp\frac{1}{2}}\zeta^{\pm\frac{3}{2}}\mathit{f}^{-\frac{1}{2}}\nn\\
&\space&\pm\beta_{\pm\frac{1}{2}}\zeta^{\pm\frac{1}{2}}\mathit{f}^{\pm\frac{1}{2}}
-3\tau_{\mp\frac{1}{2}}\zeta^{\pm\frac{3}{2}}\nu^{\mp\frac{1}{2}}
+2\tau_{\pm\frac{1}{2}}\zeta^{\pm\frac{1}{2}}\nu^{\pm\frac{1}{2}}
-3\tau_{\pm\frac{3}{2}}\zeta^{\mp\frac{1}{2}}\nu^{\pm\frac{3}{2}}
-3\tau_{\pm\frac{3}{2}}\varrho^{\mp\frac{1}{2}}\nu^{\pm\frac{3}{2}}\nn\\
&\space&+\tau_{\pm\frac{1}{2}}\varrho^{\pm\frac{1}{2}}\nu^{\pm\frac{1}{2}}
\pm \alpha_{\pm1}\kappa^0\mu^{\pm1}
\mp\alpha_0\kappa^{\pm1}\mu^0,\\
\tau_{\pm\frac{1}{2}} \delta_{\lambda}\nu^{\pm\frac{1}{2}}&=&
\partial_{t}\zeta^{\pm\frac{1}{2}}
-\chi^{\pm1}\gamma_{\pm1}\varrho^{\mp\frac{1}{2}}
-\chi^0\gamma_0\varrho^{\pm\frac{1}{2}}
\mp3\chi^{\mp1}\gamma_{\mp1}\zeta^{\pm\frac{3}{2}}
\mp\frac{1}{2}\chi^0\gamma_0\zeta^{\pm\frac{1}{2}}
\pm2\chi^{\pm1}\gamma_{\pm1}\zeta^{\mp\frac{1}{2}}\nn\\
&\space&\mp3\alpha_{\mp1}\zeta^{\pm\frac{3}{2}}\mu^{\mp1}
-\frac{1}{2}\alpha_0\zeta^{\pm\frac{1}{2}}\mu^0
\pm2\alpha_{\pm1}\zeta^{\mp\frac{1}{2}}\mu^{\pm1}
+\beta_{\pm\frac{1}{2}}\mathit{f}^{\pm\frac{1}{2}}\kappa^0
+\beta_{\mp\frac{1}{2}}\mathit{f}^{\mp\frac{1}{2}}\kappa^{\pm1}\nn\\
&\space&\pm3\tau_{\pm\frac{3}{2}}\epsilon^{\mp1}\nu^{\pm\frac{3}{2}}
\pm\frac{1}{2}\tau_{\pm\frac{1}{2}}\epsilon^0\nu^{\pm\frac{1}{2}}
\mp2\tau_{\mp\frac{1}{2}}\epsilon^{\pm1}\nu^{\mp\frac{1}{2}}
\pm\tau_{\pm\frac{3}{2}}\nu^{\pm\frac{3}{2}}\kappa^{\mp1}
\pm\frac{1}{2}\tau_{\pm\frac{1}{2}}\nu^{\pm\frac{1}{2}}\kappa^0\nn\\
&\space&\mp2\tau_{\mp\frac{1}{2}}\nu^{\mp\frac{1}{2}}\kappa^{\pm1},\\
\tau_{\pm\frac{3}{2}} \delta_{\lambda}\nu^{\pm\frac{3}{2}}&=&
 \partial_{t}\zeta^{\pm\frac{3}{2}}
-\chi^{\pm1}\gamma_{\pm1}\varrho^{\pm\frac{1}{2}}
\mp\frac{3}{2}\chi^0\gamma_0\zeta^{\pm\frac{3}{2}}
\pm\chi^{\pm1}\gamma_{\pm1}\zeta^{\pm\frac{1}{2}}
\mp\frac{3}{2}\alpha_0\zeta^{\pm\frac{3}{2}}\mu^0
+\alpha_{\pm1}\zeta^{\pm\frac{1}{2}}\mu^{\pm1}\nn\\
&\space&+\beta_{\pm\frac{1}{2}}\mathit{f}^{\pm\frac{1}{2}}\kappa^{\pm1}
\pm\frac{3}{2}\tau_{\pm\frac{3}{2}}\epsilon^0\nu^{\pm\frac{3}{2}}
-\tau_{\pm\frac{1}{2}}\epsilon^{\pm1}\nu^{\pm\frac{1}{2}}
\pm\frac{3}{2}\tau_{\pm\frac{3}{2}}\nu^{\pm\frac{3}{2}}\kappa^0
\mp\tau_{\pm\frac{1}{2}}\nu^{\pm\frac{1}{2}}\kappa^{\pm1}.
\end{eqnarray}
As in the $\mathfrak{osp}(1|2) \oplus \mathfrak{osp}(1|2)$ case one can now determine the canonical boundary charges $\mathcal{Q[\lambda]}$ that generates
the transformations $(\ref{tra211})$-$(\ref{tra217})$.\,Therefore,\,the corresponding variation of the boundary charge $\mathcal{Q[\lambda]}$ \cite{Banados:1998gg,Carlip:2005zn,Banados:1994tn,Banados:1998ta},\,to show the asymptotic symmetry algebra,\,is given by $(\ref{Qvar})$.\,The canonical boundary charge $\mathcal{Q[\lambda]}$  can be integrated which reads
\begin{equation}\label{boundaryco10}
\mathcal{Q[\lambda]}=\int\mathrm{d}\varphi\;\left[
 \mathcal{L}^{i}\epsilon^{-i}
+ \mathcal{A}^{i}\kappa^{-i}
+\mathcal{G}^{p}\zeta^{-p}
+\mathcal{S}^{p}\varrho^{-p}
\right].
\end{equation}
Having determined  both the infinitesimal transformations and the canonical boundary charges as the generators of the asymptotic symmetry algebra,\,one can also get  the Poisson bracket algebra by using the methods \cite{Blgojevic:2002} with $(\ref{Qvar2})$ again for any phase space functional $\digamma$,\,the  operator product algebra can then be defined as,
\begin{eqnarray}\label{aff2}
\mathcal{L}^{i}(z_1)\mathcal{L}^{j}(z_2)\,& \sim &\, \frac{\frac{k}{2}\eta^{ij}}{z_{12}^{2}}\,+\,\frac{(i-j)}{z_{12}} \mathcal{L}^{i+j} ,\\
\mathcal{L}^{i}(z_1)\mathcal{G}^{p}(z_2)\,& \sim &\,\frac{({i\over 2}-p)}{z_{12}} \mathcal{G}^{i+p} ,\\
\mathcal{G}^{p}(z_1)\mathcal{G}^{q}(z_2)\,& \sim &\, \frac{\frac{k}{2}\eta^{pq}}{z_{12}^{2}}\,+\,\frac{1}{z_{12}} \big(\sigma_{1}\mathcal{A}^{p+q} +\sigma_{2}\mathcal{L}^{p+q} \big),\\
\mathcal{G}^{p}(z_1)\mathcal{S}^{q}(z_2)\,& \sim &\,\frac{\big(\frac{3p}{2}-\frac{q}{2}\big)}{z_{12}} \big(\sigma_{3}\mathcal{A}^{p+q} +\sigma_{4}\mathcal{L}^{p+q} \big),\\
\mathcal{L}^{i}(z_1)\mathcal{A}^{j}(z_2)\,& \sim &\,\frac{\big(\frac{i}{2}-j\big)}{z_{12}} \mathcal{A}^{i+j} ,\\
\mathcal{L}^{i}(z_1)\mathcal{S}^{j}(z_2)\,& \sim &\,\frac{\big(\frac{3i}{2}-j\big)}{z_{12}} \mathcal{S}^{i+j} ,\\
\mathcal{A}^{i}(z_1)\mathcal{A}^{j}(z_2)\,& \sim &\,\frac{\frac{k}{2}\eta^{ij}}{z_{12}^{2}}\,+\,\frac{\big(i-j\big)}{z_{12}} \big(\sigma_{5}\mathcal{L}^{i+j} +\sigma_{6}\mathcal{A}^{i+j} \big),
\end{eqnarray}
\begin{eqnarray}
\mathcal{A}^{i}(z_1)\mathcal{G}^{p}(z_2)\,& \sim &\,\frac{1}{z_{12}}\bigg(\sigma_{7}\mathcal{S}_{i+p} +\sigma_{8}\bigg(\frac{i}{2}-p\bigg)\mathcal{G}_{i+p} \bigg),\\
\mathcal{A}^{i}(z_1)\mathcal{S}^{p}(z_2)\,& \sim &\,\frac{1}{z_{12}}\bigg(\sigma_{9}\bigg(\frac{3i}{2}-p\bigg)\mathcal{S}_{i+p}+\sigma_{10}\bigg(3 i^2-2ip+p^2-\frac{9}{4}\bigg)\mathcal{G}_{i+p} \bigg),\\
\mathcal{S}^{p}(z_1)\mathcal{S}^{q}(z_2)\,& \sim &\,\frac{\frac{k}{2}\eta^{pq}}{z_{12}^{2}}\,+\,\frac{1}{z_{12}}\bigg(\bigg(3p^2-4pq+3q^2-\frac{9}{2}\bigg)\big(\sigma_{11}\mathcal{A}_{p+q} +\sigma_{12}\mathcal{L}_{p+q} \big)\bigg)
\end{eqnarray}
where $z_{12}=z_1-z_2$,\,or in the more compact form,
\begin{eqnarray}\label{ope111}
\mathfrak{\mathcal{\mathfrak{J}}}^{A}(z_1)\mathfrak{\mathcal{\mathfrak{J}}}^{B}(z_2)\,& \sim &\, \frac{\frac{k}{2}\eta^{AB}}{z_{12}^{2}}\,+\,\frac{\mathfrak{\mathcal{\mathfrak{f}}}^{AB}_{~~~C} \mathfrak{\mathcal{\mathfrak{J}}}^{C}(z_2)}{z_{12}}.
\end{eqnarray}
Here, $\eta^{AB}$ is the supertrace matrix and $\mathfrak{\mathcal{\mathfrak{f}}}^{AB}_{~~C}$'s are the structure constants of the related algebra with $(A,B=0,\pm1,\pm\frac{1}{2},0,\pm1,\pm\frac{1}{2},\pm\frac{3}{2})$,\,i.e,\,$\eta^{ip}=0$ and $\mathfrak{\mathcal{\mathfrak{f}}}^{ij}_{~~i+j}=(i-j)$.\,After repeating the same algebra for $\bar{\Gamma}$\,-\,sector,\,one can say that the asymptotic symmetry algebra for the  most general  boundary conditions of $\mathcal{N}=(1,1)$ supergravity  is two copies of the affine $\mathfrak{osp}(3|2)_k$ algebra.
\subsection{For Superconformal Boundary}\label{scbsss}
Under the following restrictions as the Drinfeld\,-\,Sokolov highest weight gauge condition,
\begin{eqnarray}
\mathcal{L}^0&=&\mathcal{A}^0=\mathcal{A}^{+1}=\mathcal{G}^{+\frac{1}{2}}=\mathcal{S}^{+\frac{1}{2}}=\mathcal{S}^{+\frac{3}{2}}=0,\nn\\
\,\mathcal{L}^{-1}&=&\mathcal{L},\,\mathcal{A}^{-1}=\mathcal{A},\,\mathcal{G}^{-\frac{1}{2}}=\mathcal{G},\,\mathcal{S}^{-\frac{3}{2}}=\mathcal{S},\,\alpha_{+1}\mathcal{L}^{+1}=1
\end{eqnarray}
on the boundary conditions with the $\mathfrak{osp}(3|2)$ superalgebra valued connection (\ref{bouncond999}),\,one can get the superconformal boundary conditions as the supersymmetric extension of the Brown\,-\,Henneaux boundary conditions proposed in \cite{Brown:1986nw,Banados:1998pi} for $AdS_3$ supergravity.\,Therefore we have the supersymmetric connection as,
\begin{eqnarray}\label{bouncond9999}
a_\varphi &=&
 \Lt_{1}
+\alpha_{-1}\mathcal{L} \Lt_{-1}
+\gamma_{-1}\mathcal{A} \Lt_{-1}
+\beta_{-\frac{1}{2}}\mathcal{G}\Gt_{-\frac{1}{2}}
+\tau_{-\frac{3}{2}}\mathcal{S}\St_{-\frac{3}{2}}
,\\
a_t &=&
\mathcal{\mu} \Lt_{1}
+\mathcal{\chi} \At_{1}
+\mathcal{\mathit{f}}\Gt_{{+\frac{1}{2}}}
+\mathcal{\nu}\St_{{+\frac{3}{2}}}
+\sum_{i=-1}^{0}\mathcal{\mu}^i \Lt_{i}
+\sum_{i=-1}^{0}\mathcal{\chi}^i \At_{i}
+\mathcal{\mathit{f}}^{-{\frac{1}{2}}}\Gt_{-{\frac{1}{2}}}
+\sum_{p=-\frac{3}{2}}^{ \frac{1}{2}}\mathcal{\nu}^{p}\St_{p}
\end{eqnarray}
where $\mathcal{\mu}\equiv\mathcal{\mu}^{+1}$,\,
$\mathcal{\chi}\equiv\mathcal{\chi}^{+1}$,\,
$\mathcal{\nu}\equiv\mathcal{\nu}^{+{\frac{1}{2}}}$,\,and
$\mathcal{\mathcal{\mathit{f}}}\equiv\mathcal{\mathit{f}}^{+{\frac{3}{2}}}$
can be interpreted as the independent $chemical$ $potentials$.\,One can choose for the corresponding asymptotic symmetry algebra, the resulting relations as $\sigma _3\,=\, \frac{2 i}{\sqrt{5}},\,~~\sigma _7\,=\, i \sqrt{5},\,~~\sigma _8\,=\, 0,\,~~\sigma _9\,=\,1,\,$appearing on the RHS of eq.$(\ref{algebraLGASs})-(\ref{algebraLGASss})$ in the super Jacobi identities of the $\mathfrak{osp}(3|2)$ superalgebra.\,The functions,\,except the $chemical$ $potentials$  are fixed by the flatness condition (\ref{flatness}).\,For the fixed chemical potentials,\,the time evolution of canonical boundary charges can be written as,
\begin{eqnarray}
\partial_{t}\mathcal{L}	&=&
\frac{\mu '''}{2 \alpha _{-1}}+\mu  \mathcal{L}'+2 \mathcal{L}
\mu '+\frac{\gamma _{-1} \chi  \mathcal{A}'}{\alpha _{-1}}+\frac{2 \mathcal{A} \gamma _{-1} \chi '}{\alpha _{-1}}-\frac{3 \beta _{-\frac{1}{2}} \mathcal{G} \mathit{f}'}{\alpha _{-1}}\nonumber\\
&\space&-\frac{\beta   _{-\frac{1}{2}} \mathit{f} \mathcal{G}'}{\alpha _{-1}}-\frac{9 \delta _{-\frac{3}{2}} \nu  \mathcal{S}'}{10 \alpha _{-1}}-\frac{3 \delta _{-\frac{3}{2}} \mathcal{S} \nu '}{2 \alpha _{-1}},\\
\partial_{t}\mathcal{G}	&=&
\frac{3 i \gamma _{-1} \nu  \mathcal{A}'}{2 \sqrt{5} \beta _{-\frac{1}{2}}}+\frac{2 i \mathcal{A} \gamma _{-1} \nu '}{\sqrt{5} \beta _{-\frac{1}{2}}}-\frac{\mathit{f}''}{\beta _{-\frac{1}{2}}}-\frac{\alpha _{-1} \mathit{f}   \mathcal{L}}{\beta _{-\frac{1}{2}}}+\mu  \mathcal{G}'+\frac{3 \mathcal{G} \mu '}{2}+\frac{3 i \delta _{-\frac{3}{2}} \mathcal{S} \chi }{2 \sqrt{5} \beta _{-\frac{1}{2}}},\\
\partial_{t}\mathcal{A}	&=&\mu  \mathcal{A}'-\frac{5}{2} i \chi  \mathcal{A}'-\frac{9 \mathcal{A} \beta _{-\frac{1}{2}} \mathcal{G} \nu }{\sqrt{5}}+2 \mathcal{A} \mu '-5 i \mathcal{A} \chi '+\frac{\chi '''}{2 \gamma _{-1}}+\frac{3 i \delta   _{-\frac{3}{2}} \mathit{f} \mathcal{S}}{\sqrt{5} \gamma _{-1}}\nn\\
&\space&-\frac{3 i \beta _{-\frac{1}{2}} \nu  \mathcal{G}''}{2 \sqrt{5} \gamma _{-1}}
-\frac{4 i \beta _{-\frac{1}{2}} \mathcal{G}' \nu '}{\sqrt{5} \gamma _{-1}}-\frac{3 i \beta _{-\frac{1}{2}} \mathcal{G} \nu ''}{\sqrt{5} \gamma _{-1}}-\frac{9 i \alpha _{-1} \beta _{-\frac{1}{2}} \mathcal{G} \nu  \mathcal{L}}{2 \sqrt{5} \gamma _{-1}}+\frac{9 i \delta _{-\frac{3}{2}} \nu  \mathcal{S}'}{10 \gamma _{-1}}\nn\\
&\space&+\frac{3 i \delta _{-\frac{3}{2}} \mathcal{S} \nu '}{2 \gamma _{-1}}+\frac{\alpha _{-1} \chi  \mathcal{L}'}{\gamma _{-1}}+\frac{2 \alpha _{-1} \mathcal{L} \chi '}{\gamma _{-1}},
\end{eqnarray}
\begin{eqnarray}
\partial_{t}\mathcal{S}	&=&
\frac{i \gamma _{-1} \nu  \mathcal{A}''}{2 \delta _{-\frac{3}{2}}}+\frac{5 i \gamma _{-1} \mathcal{A}' \nu '}{3 \delta _{-\frac{3}{2}}}-\frac{i \sqrt{5} \gamma _{-1} \mathit{f} \mathcal{A}'}{3 \delta _{-\frac{3}{2}}}+\frac{5 i \mathcal{A} \gamma _{-1} \nu ''}{3 \delta _{-\frac{3}{2}}}-\frac{4 i \sqrt{5} \mathcal{A} \gamma _{-1} \mathit{f}'}{3 \delta _{-\frac{3}{2}}}\nn\\
&\space&-\frac{3 \sqrt{5} \mathcal{A} \beta _{-\frac{1}{2}} \gamma _{-1} \mathcal{G} \chi }{\delta _{-\frac{3}{2}}}
+\frac{3 i \alpha _{-1} \mathcal{A} \gamma _{-1} \nu  \mathcal{L}}{\delta _{-\frac{3}{2}}}-\frac{\nu ^{(4)}}{6 \delta _{-\frac{3}{2}}}-\frac{i \sqrt{5} \beta _{-\frac{1}{2}} \chi \mathcal{G}''}{6 \delta _{-\frac{3}{2}}}-\frac{7 \beta _{-\frac{1}{2}}^2 \mathcal{G} \nu  \mathcal{G}'}{2 \delta _{-\frac{3}{2}}}\nn\\
&\space&-\frac{2 i \sqrt{5} \beta _{-\frac{1}{2}} \mathcal{G}' \chi '}{3 \delta _{-\frac{3}{2}}}-\frac{i \sqrt{5} \beta _{-\frac{1}{2}} \mathcal{G} \chi ''}{\delta _{-\frac{3}{2}}}
-\frac{3 i \sqrt{5} \alpha _{-1} \beta _{-\frac{1}{2}} \mathcal{G} \chi  \mathcal{L}}{2 \delta _{-\frac{3}{2}}}+\mu \mathcal{S}'-i \chi  \mathcal{S}'+\frac{5 \mathcal{S} \mu '}{2}\nn\\
&\space&-\frac{5}{2} i \mathcal{S} \chi '-\frac{\alpha _{-1} \nu  \mathcal{L}''}{2 \delta _{-\frac{3}{2}}}-\frac{3 \alpha _{-1}^2 \nu  \mathcal{L}^2}{2 \delta _{-\frac{3}{2}}}-\frac{5 \alpha _{-1} \nu ' \mathcal{L}'}{3 \delta _{-\frac{3}{2}}}-\frac{5 \alpha _{-1} \mathcal{L} \nu ''}{3 \delta _{-\frac{3}{2}}}
\end{eqnarray}
where $\alpha_{-1}$,\,$\gamma_{-1}$,\,$\beta_{-{\frac{1}{2}}}$ and $\tau_{-{\frac{3}{2}}}$ are some scaling parameters to be determined later.\,Now,\,we are in a position to work  the superconformal asymptotic symmetry algebra under the Drinfeld\,-\,Sokolov reduction.\,This reduction implies that the only independent parameters
$\mathcal{\epsilon}\equiv\mathcal{\epsilon}^{+1}$,\,
$\mathcal{\kappa}\equiv\mathcal{\kappa}^{+1}$,\,
$\mathcal{\zeta}\equiv\mathcal{\zeta}^{+{\frac{1}{2}}}$ and
$\mathcal{\varrho}\equiv\mathcal{\varrho}^{+{\frac{3}{2}}}$
.\,One can start to compute the gauge transformations for asymptotic symmetry algebra by considering all transformations $(\ref{boundarycond})$ that preserve the boundary conditions with the gauge parameter $\lambda$ in the $\mathfrak{osp}(3|2)$ superalgebra $(\ref{algebraLGASs})-(\ref{algebraLGASss})$ as
\begin{eqnarray}
 \lambda\,&=&\,b^{-1}\bigg[\mathcal{
 \epsilon} \Lt_{+1}
-\epsilon' \Lt_{0}
+\bigg(
\mathcal{A} \gamma _{-1} \kappa -\beta _{-\frac{1}{2}} \zeta  \mathcal{G}-\frac{9}{10} \delta _{-\frac{3}{2}} \rho  \mathcal{S}+\alpha _{-1} \mathcal{L} \epsilon +\frac{\epsilon ''}{2}
 \bigg)\Lt_{-1}\nn\\
&\space&+\zeta\Gt_{+\frac{1}{2}} +\bigg(
\frac{3 i  \gamma _{-1} \varrho \mathcal{A} }{2 \sqrt{5}}-\zeta '+\beta _{-\frac{1}{2}} \mathcal{G} \epsilon
\bigg)\Gt_{-\frac{1}{2}}+\kappa \At_{+1}+ \left(-\kappa '+\frac{3 i \beta _{-\frac{1}{2}} \varrho \mathcal{G} }{\sqrt{5}}\right)\At_{0}\nn\\
&\space&+\left(-\frac{5}{2} i \mathcal{A} \gamma _{-1} \kappa +\mathcal{A} \gamma _{-1} \epsilon -\frac{3 i \beta _{-\frac{1}{2}} \rho  \mathcal{G}'}{2 \sqrt{5}}-\frac{1}{2} i \sqrt{5} \beta _{-\frac{1}{2}} \mathcal{G} \varrho   '+\frac{\kappa ''}{2}+\frac{9}{10} i \delta _{-\frac{3}{2}} \rho  \mathcal{S}+\alpha _{-1} \kappa  \mathcal{L}\right) \At_{-1}\nn\\
&\space&+\varrho  \St_{\frac{3}{2}}-  \varrho '\St_{\frac{1}{2}}+ \left(-\frac{3}{2} i \mathcal{A} \gamma _{-1} \rho +\frac{1}{2} i \sqrt{5} \beta _{-\frac{1}{2}} \mathcal{G} \kappa +\frac{\varrho ''}{2}+\frac{3}{2} \alpha _{-1} \varrho  \mathcal{L}\right)\St_{-\frac{1}{2}}\nn\\
&\space&+ \bigg(\frac{1}{2} i \gamma _{-1} \varrho  \mathcal{A}'-\frac{1}{3} i \sqrt{5} \mathcal{A} \gamma _{-1} \zeta +\frac{7}{6} i \mathcal{A} \gamma _{-1} \varrho '+\beta _{-\frac{1}{2}}^2 -\mathcal{G}^2 \varrho -\frac{1}{6} i \sqrt{5} \beta _{-\frac{1}{2}} \kappa  \mathcal{G}'\nn\\
&\space&-\frac{1}{2} i \sqrt{5} \beta _{-\frac{1}{2}} \mathcal{G} \kappa '
-\frac{\varrho ^{(3)}}{6}-i \delta _{-\frac{3}{2}} \kappa  \mathcal{S}+\delta _{-\frac{3}{2}}\mathcal{S} \epsilon -\frac{1}{2} \alpha _{-1} \varrho  \mathcal{L}'-\frac{7}{6} \alpha _{-1} \mathcal{L} \varrho '\bigg) \St_{-\frac{3}{2}}
 \bigg]b.
\end{eqnarray}
The condition $(\ref{boundarycond})$ impose that transformations on the gauge with $\alpha_{-1}=\gamma_{-1}=\frac{6}{c}$ ,\,$\beta_{-{\frac{1}{2}}}=-\frac{3}{c}$ and $\tau_{-{\frac{3}{2}}}=-\frac{10}{c}$ are given by
\begin{eqnarray}
\delta_{\lambda}\mathcal{L}	&=&\frac{c \epsilon'''}{12}+\epsilon  \mathcal{L}'+2
   \mathcal{L} \epsilon ' +\kappa  \mathcal{A}'+2 \mathcal{A} \kappa '  +\frac{\zeta  \mathcal{G}'}{2}+\frac{3 \mathcal{G} \zeta '}{2}+\frac{3 \rho  \mathcal{S}'}{2}+\frac{5 \mathcal{S} \rho '}{2}, \\
\delta_{\lambda}\mathcal{G}	&=&-\frac{3 i \rho  \mathcal{A}'}{\sqrt{5}}-\frac{4 i \mathcal{A} \rho '}{\sqrt{5}}+\frac{c \zeta ''}{3}+\epsilon  \mathcal{G}'+\frac{3 \mathcal{G} \epsilon '}{2}+i \sqrt{5} \kappa  \mathcal{S}+2 \zeta  \mathcal{L},\\
\delta_{\lambda}\mathcal{A}	&=&-\frac{5}{2} i \kappa  \mathcal{A}'+\epsilon  \mathcal{A}'-5 i \mathcal{A} \kappa '+2 \mathcal{A} \epsilon '+\frac{27 \mathcal{A} \mathcal{G} \rho }{\sqrt{5} c}+\frac{27 i \mathcal{G} \rho  \mathcal{L}}{2 \sqrt{5}   c}+\frac{c \kappa '''}{12}  \nn\\
&\space&+\frac{3 i \rho  \mathcal{G}''}{4 \sqrt{5}}+\frac{2 i \mathcal{G}' \rho '}{\sqrt{5}}+\frac{3 i \mathcal{G} \rho ''}{2 \sqrt{5}}-\frac{3}{2} i \rho  \mathcal{S}'-i \sqrt{5} \zeta\mathcal{S}-\frac{5}{2} i \mathcal{S} \rho '+\kappa  \mathcal{L}'+2 \mathcal{L} \kappa ',
\end{eqnarray}
\begin{eqnarray}
\delta_{\lambda}\mathcal{S}	&=&-\frac{3}{10} i \rho  \mathcal{A}''+\frac{i \zeta  \mathcal{A}'}{\sqrt{5}}-i \mathcal{A}' \rho '+\frac{4 i \mathcal{A} \zeta '}{\sqrt{5}}-i \mathcal{A} \rho ''-\frac{27 \mathcal{A} \mathcal{G} \kappa }{\sqrt{5} c}-\frac{54 i\mathcal{A} \rho  \mathcal{L}}{5 c}+\frac{63 \mathcal{G} \rho  \mathcal{G}'}{20 c}\nn\\
&\space&-\frac{27 i \mathcal{G} \kappa  \mathcal{L}}{2 \sqrt{5} c}+\frac{c \rho ^{(4)}}{60}+\frac{27 \rho  \mathcal{L}^2}{5 c}-\frac{i \kappa   \mathcal{G}''}{4 \sqrt{5}}-\frac{i \mathcal{G}' \kappa '}{\sqrt{5}}-\frac{3 i \mathcal{G} \kappa ''}{2 \sqrt{5}}-i \kappa  \mathcal{S}'+\epsilon  \mathcal{S}'-\frac{5}{2} i \mathcal{S} \kappa '\nn\\
&\space&+\frac{5 \mathcal{S} \epsilon   '}{2}+\frac{3 \rho  \mathcal{L}''}{10}+\rho ' \mathcal{L}'+\mathcal{L} \rho ''.
\end{eqnarray}
The variation of  canonical boundary charge $\mathcal{Q[\lambda]}$ $(\ref{Qvar})$  can be integrated which reads
\begin{equation}
\mathcal{Q[\lambda]}=\int\mathrm{d}\varphi\;\left[\mathcal{L}\epsilon+\mathcal{A}\kappa+\mathcal{G}\zeta+\mathcal{S}\varrho\right].\label{boundaryco11111}
\end{equation}
This leads to operator product expansions in the complex coordinates by using $(\ref{Qvar2})$
\begin{subequations}
\begin{align}
&\mathcal{L}(z_1)\mathcal{L}(z_2)\,\sim  \,{{c\over 2}\over{z_{12}^{4}}}\,+\, {2\,\mathcal{L}\over{z_{12}^{2}}}\, + \,{ \mathcal{L}'\over{z_{12}}},\\
&\mathcal{L}(z_1)\mathcal{G}(z_2)\, \sim  \,{\frac{3}{2}\mathcal{G}\over{z_{12}^{2}}}\, + \,{ \mathcal{G}'\over{z_{12}}},~~~
 \mathcal{G}(z_1)\mathcal{G}(z_2)\,\sim \,{{2c\over 3}\over{z_{12}^{3}}}\,+\, {2\,\mathcal{L}\over{z_{12}}},~\\
&\mathcal{L}(z_1)\mathcal{A}(z_2)\, \sim  \,{2 \mathcal{A}\over{z_{12}^{2}}}\, + \,{ \mathcal{A}'\over{z_{12}}},~~~
 \mathcal{L}(z_1)\mathcal{S}(z_2)\, \sim  \,{\frac{5}{2}\mathcal{S}\over{z_{12}^{2}}}\, + \,{ \mathcal{S}'\over{z_{12}}},~~~\\
&\mathcal{G}(z_1)\mathcal{A}(z_2)\,\sim \,- {i\sqrt{5}\,\mathcal{A}\over{z_{12}}},~~~\mathcal{G}(z_1)\mathcal{S}(z_2)\,\sim \,\,{\frac{4i}{\sqrt{5}}\mathcal{A}\over{z_{12}^{2}}}\, -{{\frac{i}{\sqrt{5}}\mathcal{A}'}\over{z_{12}}},\\
&\mathcal{A}(z_1)\mathcal{A}(z_2)\,\sim  \,{{c\over 2}\over{z_{12}^{4}}}\,+\, {{2\,\mathcal{L}-5i\mathcal{A}}\over{z_{12}^{2}}}\, + \,{{\,\mathcal{L}-\frac{5i}{2}\mathcal{A}}\over{z_{12}}},\\
&\mathcal{A}(z_1)\mathcal{S}(z_2)\,\sim  \,
-\,{\frac{3i}{\sqrt{5}}\mathcal{G}\over{z_{12}^{3}}}
\,-\,\frac{\frac{i\mathcal{G}'}{\sqrt{5}}+\frac{5 i \mathcal{S}}{2}}{z_{12}^2}
\,-\,\frac{1}{z_{12}}\bigg(
{\frac{27 \mathcal{A} \mathcal{G}}{\sqrt{5} c}+\frac{27 i \mathcal{G} \mathcal{L}}{2 \sqrt{5} c}+i \mathcal{S}'+\frac{i \mathcal{G}''}{4 \sqrt{5}} }
\bigg), \\
&\mathcal{S}(z_1)\mathcal{S}(z_2)\,\sim  \,
 {{{2c}\over 5}\over{z_{12}^{5}}}
 +\frac{2 (\mathcal{L}-i \mathcal{A})}{z_{12}^3}
 +\frac{\mathcal{L}'-i \mathcal{A}'}{z_{12}^2}
 +\frac{1}{z_{12}}\bigg({\frac{27 \mathcal{L}^2}{5 c}-\frac{54 i \mathcal{A} \mathcal{L}}{5 c}+\frac{63 \mathcal{G} \mathcal{G}'}{20 c}-\frac{3 i \mathcal{A}''}{10} +\frac{3
   \mathcal{L}''}{10}}\bigg).
\end{align}
\end{subequations}
After repeating the same algebra for $\bar{\Gamma}$\,-\,sector,\,one can say that the asymptotic symmetry algebra for a set of boundary conditions of $\mathcal{N}=(1,1)$ supergravity  is two copies of the $\mathcal{SW}(\frac{3}{2},2)$ algebra  with central charce $c\,=\,6k$.\,Finally,\,one can also take into account normal ordering (quantum) effects of  this algebra as in Ref's.\,\cite{Figueroa-OFarrill:1996tnk,Figueroa-OFarrill:1990tqt}.
\section{Other extended (super)gravity checks in the most general boundary conditions\,}\label{sec2}

In this section,\,it will be very interesting to show  another class of boundary conditions that appeared in the literature (see,\,eg \cite{Compere:2013bya,Afshar:2016wfy,Troessaert:2013fma,Avery:2013dja}) for gravity case,\,whose higher\,-\,spin generalization is less
clear than the Grumiller\,-\,Riegler boundary conditions.\,Because of the possibility to obtain asymptotic symmetries that are the affine
version of the gauge algebras of the Chern-Simons theory is not a surprise given the relation between Chern-Simons theories and Wess-Zumino-Witten models\,\cite{Elitzur:1989nr}.\,Therefore,\,we present two most general boundary conditions for pure bosonic gravity and $\mathcal{N}=1$ extended
supergravity.\,That is,\,The Avery\,-\,Poojary\,-\,Suryanarayana gravity for $\mathfrak{sl}(2,\mathbb{R})\oplus\mathfrak{sl}(2)$
and $\mathfrak{osp}(1|2)\oplus\mathfrak{sl}(2)$ respectively.\,In order to check,\,we compare the most general boundary conditions used so far
with the Avery\,-\,Poojary\,-\,Suryanarayana boundary conditions for $\mathfrak{sl}(2,\mathbb{R})\oplus\mathfrak{sl}(2)$  pure gravity in section
\ref{sec22} and $\mathcal{N}=1$ supergravity for $\mathfrak{osp}(1|2)\oplus\mathfrak{sl}(2)$ in section \ref{sec2} respectively after we have worked
with the Brown\,-\,Henneaux boundary conditions for $\mathcal{N}=1$ supergravity for $\mathfrak{osp}(1|2)\oplus\mathfrak{sl}(2)$ in section \ref{scbsss}.
\,We also mention the checks that our boundary conditions passed and discuss conservation of the charges and consistency of the variational principle
$\mathcal{N}=(1,1)$ supergravity for $\mathfrak{osp}(1|2)$ in section \ref{osp12} and  $\mathfrak{osp}(3|2)$  in section \ref{osp32} respectively.

\subsection{Avery\,-\,Poojary\,-\,Suryanarayana for $\mathfrak{sl}(2,\mathbb{R})\oplus\mathfrak{sl}(2)$}\label{sec22}
The Avery\,-\,Poojary\,-\,Suryanarayana boundary conditions proposed in \cite{Avery:2013dja} are a generalization of the Brown\,-\,Henneaux boundary conditions
and that can be obtained by choosing for $\Gamma$\,-\,sector the Brown\,-\,Henneaux boundary conditions while leaving all charges to vary for the
$\bar{\Gamma}$\,-\,sector.\,Consequently,\,these lead to an interesting set of asymptotic symmetries in the form of
$\mathfrak{sl}(2,\mathbb{R})\oplus\mathfrak{sl}(2)_k$.\,To this end,\,we recapitulate  Avery\,-\,Poojary\,-\,Suryanarayana gravity for $\mathfrak{sl}(2,\mathbb{R})\oplus\mathfrak{sl}(2)$ from the most general Grumiller\,-\,Riegler gravity for
$\mathfrak{sl}(2,\mathbb{R})\oplus\mathfrak{sl}(2,\mathbb{R})$ as in \cite{Grumiller:2016pqb}.\,Therefore,\,the most general connections for Grumiller\,-\,Riegler
boundary conditions  is given by
\begin{eqnarray}\label{bouncond9999}
&&\Gamma-\textrm{sector}:~~~~
a_\varphi = \alpha_{i}\mathcal{L}^{i} \Lt_{i},~~
a_t = {\mu}^{i} \Lt_{i},\\
&&\bar{\Gamma}-\textrm{sector}:~~~~
\bar{a}_\varphi = -\alpha_{i}\bar{\mathcal{L}}^{i} \Lt_{i},~~
a_t = {\bar{\mu}}^{i} \Lt_{i}.
\end{eqnarray}
Besides,\,in the Avery\,-\,Poojary\,-\,Suryanarayana gravity the boundary conditions  is given by
\begin{eqnarray}
&&\Gamma-\textrm{sector}:~~
a_\varphi = \Lt_{1}-\kappa \Lt_{-1},\;\;\;\;
a_{t} = \Lt_{1}-\kappa \Lt_{-1},\label{r1a}\\
&&\bar{\Gamma}-\textrm{sector}:~~
\bar{a}_\varphi = \Lt_{-1}-\bar{\kappa} \Lt_{1}+\mathit{f}^{i}\Lt_{i},\;\;\;\;
\bar{a}_{t} =-\Lt_{-1}+\bar{\kappa} \Lt_{1}+\mathit{f}^{i}\Lt_{i}.\label{r1b}
\end{eqnarray}
In this gravity language, this amounts to the following restrictions on the charge and the chemical potential:
\begin{eqnarray}
&&\Gamma-\textrm{sector}:~~
\mathcal{L}^{0}=0,\;\;\;\;
\mathcal{L}^{-1}=\mathcal{L}=-\frac{\kappa}{\alpha_{-1}},\;\;\;\;
\mathcal{L}^{1}=\frac{1}{\alpha_{1}},\label{q3a}\\
&&~~~~~~~~~~~~~~~~~~~
\mu^0=0,\;\;\;\;
\mu^1=\mu=1,\;\;\;\;
\mu^{-1}=\kappa,\label{q3b}\\
&&\bar{\Gamma}-\textrm{sector}:~~
\bar{\mathcal{L}}^{0}=-\frac{\mathit{f}^{0}}{\alpha_{0}},\;\;\;\;
\bar{\mathcal{L}}^{-1}=-\frac{1}{\alpha_{-1}}( {1+\mathit{f}^{-1}}),\;\;\;\;
\bar{\mathcal{L}}^{1}=\frac{1}{\alpha_{1}}( {\bar{\kappa}-\mathit{f}^{1}}),\label{q3c}\\
&&~~~~~~~~~~~~~~~~~~~
\bar{\mu}^0=\mathit{f}^{0},\;\;\;\;
\bar{\mu}^1=\mu=\bar{\kappa}+\mathit{f}^{1},\;\;\;\;
\bar{\mu}^{-1}=\mathit{f}^{-1}-1,\,\textrm{and}\,\,\,\bar{\mathcal{L}}^{a}=\mathcal{T}^{a} .\label{q3b}
\end{eqnarray}
Next we want to find two sets of gauge transformations that preserve the boundary conditions.\,First,\,we define the gauge
parameter $\lambda$ for $\Gamma$\,-\,sector as:
\begin{eqnarray}
a_\varphi = \Lt_{1}-\alpha_{-1}\mathcal{L} \Lt_{-1},\;\;\;\;
\lambda = {\epsilon}^{i} \Lt_{i},\label{r1a}
\end{eqnarray}
where $\alpha_{-1}=\frac{6}{c}$.\,If $\Lt_{i},(i=\pm1,0)$ are the generators of $\mathfrak{sl}(2,\mathbb{R})$ algebra such that $\left[\Lt_{i},\Lt_{j}\right]= \left(i-j\right)\Lt_{i+j}$
that is,\,this is the Brown\,-\,Henneaux boundary condition that produce the following asymptotic symmetry algebra :
\begin{eqnarray}\label{ope}
&\mathcal{L}(z_1)\mathcal{L}(z_2)\,\sim  \,{{c\over 2}\over{z_{12}^{4}}}\,+\, {2\,\mathcal{L}\over{z_{12}^{2}}}\, + \,{\partial \mathcal{L}\over{z_{12}}}
\end{eqnarray}
with central charge $c=6k$.\,Second,\,we define the gauge parameter $\bar{\lambda}$ for $\bar{\Gamma}$\,-\,sector as:
\begin{eqnarray}
\bar{a}_\varphi = \Lt_{-1}-\alpha_{-1}\mathcal{L} \Lt_{1}+\beta_{i}\mathcal{T}^{i}\Tt_{0}^{i},\;\;\;\;
\bar{\lambda} = {\eta}^{i} \Lt_{i}+{\sigma}^{i}\Tt_{0}^{i}.\label{r1a}
\end{eqnarray}
where $\alpha_{-1}=\frac{6}{c},\,\beta_{-1}=\beta_{1}=-\frac{2}{k},\,\textrm{and}\,\beta_{0}=\frac{4}{k}$.\,If $\Tt_{n}^{a},\,(a=\pm1,0)$  are the generators of $\mathfrak{sl}(2)$ current algebra.\,We have expressed $\mathfrak{sl}(2,\mathbb{R})\oplus\mathfrak{sl}(2)$ algebra such that
\be
\label{algebraLG}
\left[\Lt_{n},\Lt_{m}\right]= \left(n-m\right)\Lt_{n+m},\;\qquad\left[\Lt_{n},\Tt_{m}^{a}\right]=-m\Tt_{n+m}^{a},\;\qquad[\Tt_{n}^{a},\Tt_{m}^{b}] =(a-b)\Tt_{n+m}^{a+b} .
\ee
that is,\,this is an affine boundary condition that produce the following $\mathfrak{sl}(2,\mathbb{R})\oplus\mathfrak{sl}(2)_k$ asymptotic symmetry algebra :
\begin{eqnarray}\label{ope}
&\mathcal{L}(z_1)\mathcal{L}(z_2)\,\sim  \,{{c\over 2}\over{z_{12}^{4}}}\,+\, {2\,\mathcal{L}\over{z_{12}^{2}}}\, + \,{ \mathcal{L}^{'}\over{z_{12}}},\\
&\mathcal{L}(z_1)\mathcal{T}^{a}(z_2)\, \sim  \,{ \mathcal{T}^{a}\over{z_{12}^{2}}}\, + \,{ \mathcal{T}^{{a}{'}}\over{z_{12}}},~~~
\mathcal{T}^{a}(z_1)\mathcal{T}^{b}(z_2)\, \sim \, \frac{\frac{k}{2}\eta^{ab}}{z_{12}^{2}}\,+\,\frac{(a-b)}{z_{12}} \mathcal{T}^{a+b}(z_2).
\end{eqnarray}
with central charge $c=6k$
and performing a Sugawara shift \cite{Grumiller:2016pqb},\,
\begin{equation}
    \mathcal{L}\rightarrow\mathcal{L}^{-1}+\frac{1}{k}\left(\mathcal{T}^{0}\mathcal{T}^{0}-\mathcal{T}^{+}\mathcal{T}^{-}\right).\label{shift01}
\end{equation}
\subsection{Avery\,-\,Poojary\,-\,Suryanarayana  $\mathcal{N}=1$ supergravity for $\mathfrak{osp}(1|2)\oplus\mathfrak{sl}(2)$}\label{sec2}
Inspired by previous calculations,\,in this section,\,we calculate  Avery\,-\,Poojary\,-\,Suryanarayana $\mathcal{N}=1$ gravity for $\mathfrak{osp}(1|2)\oplus\mathfrak{sl}(2)$ from the most general Grumiller\,-\,Riegler $\mathcal{N}=(1,1)$ supergravity for $\mathfrak{osp}(1|2)\oplus\mathfrak{osp}(1|2)$.\,Therefore,\,the most general connections for Grumiller\,-\,Riegler boundary conditions  is given by

\begin{eqnarray}\label{bouncond9999}
\Gamma-\textrm{sector}:~~~~
a_\varphi &=&
 \alpha_i\mathcal{L}^i \Lt_{i}
+\beta_p\mathcal{G}^{p}\Gt_{p}
,~~\\
a_t &=&
 \mu^i \Lt_{i}
+\nu^{p}\Gt_{p}
,\\
\bar{\Gamma}-\textrm{sector}:~~~~
\bar{a}_\varphi &=&-(
 \alpha_i\bar{\mathcal{L}}^i \Lt_{i}
+\beta_p\bar{\mathcal{G}}^{p}\Gt_{p})
,~~\\
\bar{a}_t &=&
 \bar{\mu}^i \Lt_{i}
+\bar{\nu}^{p}\Gt_{p}.
\end{eqnarray}
Besides,\,in the Avery\,-\,Poojary\,-\,Suryanarayana gravity the boundary conditions  is given by
\begin{eqnarray}
\Gamma-\textrm{sector}:~~
a_\varphi &=& \Lt_{1}-\kappa \Lt_{-1}-\omega \Gt_{-\frac{1}{2}},\;\;\;\;\\
a_{t} &=& \Lt_{1}-\kappa \Lt_{-1}-\omega \Gt_{-\frac{1}{2}},\label{r1a}\\
\bar{\Gamma}-\textrm{sector}:~~
\bar{a}_\varphi &=& \Lt_{-1}-\bar{\kappa} \Lt_{1}-\bar{\omega} \Gt_{\frac{1}{2}}+\mathit{f}^{i}\Lt_{i}+\mathit{g}^{p}\Gt_{p},\;\;\;\;\\
\bar{a}_{t} &=&-\Lt_{-1}+\bar{\kappa} \Lt_{1}+\bar{\omega} \Gt_{\frac{1}{2}}+\mathit{f}^{i}\Lt_{i}+\mathit{g}^{p}\Gt_{p}.\label{r1b}
\end{eqnarray}
In this gravity language, this amounts to the following restrictions on the charges and the chemical potentials:
\begin{eqnarray}
&&\Gamma-\textrm{sector}:~~
\mathcal{L}^{0}=0,\;
\mathcal{L}^{-1}=\mathcal{L}=-\frac{\kappa}{\alpha_{-1}},\;
\mathcal{L}^{1}=\frac{1}{\alpha_{1}},\;
\mathcal{G}^{-\frac{1}{2}}=-\frac{\omega}{\beta_{-\frac{1}{2}}},\;
\mathcal{G}^{\frac{1}{2}}=0,\label{q3a}
\\
&&~~~~~~~~~~~~~~~~~~~
\mu^0=0,\;\;
\mu^1=\mu=1,\;
\mu^{-1}=\kappa,\;
\nu^{-\frac{1}{2}}=-\omega,\;
\nu^{ \frac{1}{2}}=0,\;
\label{q3b}
\\
&&\bar{\Gamma}-\textrm{sector}:~~
\bar{\mathcal{L}}^{0}=-\frac{\mathit{f}^{0}}{\alpha_{0}},\;
\bar{\mathcal{L}}^{-1}=-\frac{1}{\alpha_{-1}}( {1+\mathit{f}^{-1}}),\;
\bar{\mathcal{L}}^{1}=\frac{1}{\alpha_{1}}( {\bar{\kappa}-\mathit{f}^{1}}),\;
\label{q3c}\\
&&~~~~~~~~~~~~~~~~~~~
\bar{\mathcal{G}}^{-\frac{1}{2}}=\frac{g^{-\frac{1}{2}}}{\beta_{-\frac{1}{2}}},\;
\bar{\mathcal{G}}^{\frac{1}{2}}=\frac{\bar{\omega}-g^{\frac{1}{2}}}{\beta_{\frac{1}{2}}},\;
\bar{\mu}^0=\mathit{f}^{0},\;\;\;\;
\bar{\mu}^1=\mu=\bar{\kappa}+\mathit{f}^{1},\;\;\;\;\\
&&~~~~~~~~~~~~~~~~~~~
\bar{\mu}^{-1}=\mathit{f}^{-1}-1,\;
\bar{\nu}^{-\frac{1}{2}}=g^{-\frac{1}{2}},\;
\bar{\nu}^{ \frac{1}{2}}=\bar{\omega}+g^{\frac{1}{2}},\;
\,\textrm{and}\,\,\,\bar{\mathcal{L}}^{a}=\mathcal{T}^{a} .\label{q3b}
\end{eqnarray}
Next,\,we want to find two sets of gauge transformations that preserve the boundary conditions.\,First,\,we define the gauge
parameter $\lambda$ for $\Gamma$\,-\,sector as:
\begin{eqnarray}
a_\varphi &=& \Lt_{1}-\alpha_{-1}\mathcal{L} \Lt_{-1}-\beta_{-\frac{1}{2}}\mathcal{G}\Gt_{-\frac{1}{2}},\;\;\;\;
\lambda = {\epsilon}^{i} \Lt_{i}+\zeta^{p} \Gt_{p}.\label{r1a}
\end{eqnarray}
This is the Brown\,-\,Henneaux boundary condition for the $\Gamma-\textrm{sector}$ that produces one copy of the asymptotic symmetry algebra
as in section $\ref{bhreduction1}$.\,Second,\,we define the gauge parameter $\bar{\lambda}$ for $\bar{\Gamma}$\,-\,sector as:
\begin{eqnarray}
\bar{a}_\varphi = \Lt_{-1}-\alpha_{-1}\mathcal{L} \Lt_{1}-\beta_{-\frac{1}{2}}\mathcal{G}\Gt_{\frac{1}{2}}+\gamma_{i}\mathcal{T}^{i}\Tt_{0}^{i},\;\;\;\;
\bar{\lambda} = {\epsilon}^{i} \Lt_{i}+\zeta^{p} \Gt_{p}+{\sigma}^{i}\Tt_{0}^{i}.\label{r1a}
\end{eqnarray}
where $\alpha_{-1}=\frac{6}{c},\,\beta_{-\frac{1}{2}}=-\frac{3}{c},\,\gamma_{-1}=\gamma_{1}=-\frac{2}{k},\,\textrm{and}\,\gamma_{0}=\frac{4}{k}$.\,If $\Tt_{n}^{a},\,(a=\pm1,0)$  are the generators of $\mathfrak{sl}(2)$  current algebra.\,We have expressed $\mathfrak{osp}(1|2)\oplus\mathfrak{sl}(2)$ superalgebra such that
\begin{eqnarray}\label{comm}
&&\left[\Lt_{n},\Lt_{m}\right]= \left(n-m\right)\Lt_{n+m},\;
\left[\Lt_{n},\Gt_{p}\right]=\big(\frac{n}{2}-p\big)\Gt_{n+p},\;\\
\left\{\Gt_{p},\Gt_{q}\right\}&&=-2\,\Lt_{p+q},\;
\left[\Lt_{n},\Tt_{m}^{a}\right]=-m\Tt_{n+m}^{a},\;
\left[\Tt_{n}^{a},\Tt_{m}^{b}\right] =(a-b)\Tt_{n+m}^{a+b}.
\end{eqnarray}
that is,\,this is an affine boundary condition that produces the following $\mathfrak{osp}(1|2)\oplus\mathfrak{sl}(2)_k$ asymptotic symmetry algebra :
\begin{eqnarray}\label{ope}
&\mathcal{L}(z_1)\mathcal{L}(z_2)\,\sim  \,{{c\over 2}\over{z_{12}^{4}}}\,+\, {2\,\mathcal{L}\over{z_{12}^{2}}}\, + \,{  \mathcal{L}^{'}\over{z_{12}}},\\
&\mathcal{L}(z_1)\mathcal{G}(z_2)\, \sim  \,{\frac{3}{2}\mathcal{G}\over{z_{12}^{2}}}\, + \,{ \mathcal{G}'\over{z_{12}}},~~~
\mathcal{G}(z_1)\mathcal{G}(z_2)\,\sim \,{{2c\over 3}\over{z_{12}^{3}}}\,+\, {2\,\mathcal{L}\over{z_{12}}},~~~\\
&\mathcal{L}(z_1)\mathcal{T}^{a}(z_2)\, \sim  \,{ \mathcal{T}^{a}\over{z_{12}^{2}}}\, + \,{  \mathcal{T}^{{a}{'}}\over{z_{12}}},~~~
\mathcal{T}^{a}(z_1)\mathcal{T}^{b}(z_2)\, \sim \, \frac{\frac{k}{2}\eta^{ab}}{z_{12}^{2}}\,+\,\frac{(a-b)}{z_{12}} \mathcal{T}^{a+b}.
\end{eqnarray}
with central charge $c=6k$,\, and the same Sugawara shift $(\ref{shift01})$.
\subsection{The on\,-\,shell action and the variational principle}
It is well known that on manifold $\mathcal{M}$ with boundary $\partial\mathcal{M}$ the Chern\,-\,Simons action ($\ref{CSA}$) in general is neither differentiable nor gauge invariant.\,
Therefore,\,it is important to show that the boundary conditions admit a well-defined variational principle.\,To see this,\,take a closer look at the variation of the Chern\,-\,Simons action,
\begin{equation}
\delta S_{\textrm{\tiny CS}}[\Gamma] = \frac{k}{2\pi}\int_{\mathcal{M}}\,\langle\delta\Gamma\wedge F\rangle\,+\,\frac{k}{4\pi}\int_{\mathcal{\partial M}}\,\langle\delta\Gamma\wedge \Gamma\rangle\label{CSAA},
\end{equation}
where,\,due that on\,-\,shell $F\,=\,0$,\,the on\,-\,shell bulk part is zero and we are left with only a boundary term\,:
\begin{equation}
\delta S_{\textrm{\tiny CS}}[\Gamma] = \,\frac{k}{4\pi}\int_{\mathcal{\partial M}}\extd t\extd\varphi\,\langle\Gamma_{\varphi}\wedge \delta\Gamma_{t}-\Gamma_{t}\wedge\delta \Gamma_{\varphi}\rangle\label{CSAB}.
\end{equation}
Therefore,\,it would be good to have an appropriate way of extending the possible consistent boundary conditions.\,A consistent way is to add a boundary term $ S_{\textrm{\tiny B }}[\Gamma]$ to the Chern\,-\,Simons action.\,So,\,the total action will then be of the form\,:
\begin{equation}\label{totalaction}
 S_{\textrm{\tiny tot}}\,\, =\,\,S_{\textrm{\tiny CS}}\,+\,\frac{k}{4\pi}\int_{\mathcal{\partial M}}\extd t\extd\varphi\,\langle\Gamma_{t}\wedge \Gamma_{\varphi}\rangle.
\end{equation}
\subsection{$\mathfrak{osp}(1|2)$\,-\,case\,:}\label{osp12}
Then,\,considering a solution of the form ($\ref{bouncond9999}$),\,for our boundary conditions is very easy to see that the radial dependent group element plays no role in this analysis,\,in fact plugging ($\ref{bouncond9999}$) into ($\ref{totalaction}$) we have\,:
\begin{eqnarray}
\delta S_{\textrm{\tiny tot}}[\Gamma] &=&-\frac{k}{2\pi}\,\int_{\partial\mathcal M}\!\!\!\extd t\extd\varphi\,\langle a_\varphi \,\delta a_t \rangle\nonumber\\
&=&\frac{k}{2\pi}\,\int_{\partial\mathcal M}\!\!\!\extd t\extd\varphi\,\langle\mathcal{L}^i\delta\mu^{-i}+\mathcal{G}^{p}\delta\nu^{-p}\rangle\nonumber\\
&=&0
\end{eqnarray}
because of $\delta a_t=0$.\,A similar calculation should be made for the barred sector.\,This solution ensures for that $\alpha_1=-\frac{2 \pi}{k},\,\alpha_0=\frac{4 \pi}{k},\,\alpha_{-1}=-\frac{2 \pi}{k},\,\beta_{\frac{1}{2}}=\frac{\pi}{k}$ and $\beta_{-\frac{1}{2}}=-\frac{\pi}{k}$ in ($\ref{bouncond999}$).
\subsection{$\mathfrak{osp}(3|2)$\,-\,case\,:}\label{osp32}
After the calculation of $\mathfrak{osp}(1|2)$\,-\,case,\,it could be an interesting straightforward exercise to establish a well-defined variational principle also for $\mathcal{N}=(1,1)$ extended higher\,-\,spin supergravity with two copies of the $\mathfrak{osp}(3|2)_k$ affine algebra as follows.\,To this end,\,
considering a solution of the form ($\ref{bouncondosp32}$),\,plugging these into ($\ref{totalaction}$) we have\,:
\begin{eqnarray}
\delta S_{\textrm{\tiny tot}}[\Gamma] &=&\frac{k}{2\pi}\,\int_{\partial\mathcal M}\!\!\!\extd t\extd\varphi\,\langle
\tilde{\mathcal{L}}^i\delta\mu^{-i}+\tilde{\mathcal{A}}^i\delta\chi^{-i}+\mathcal{G}^{p}\delta\mathit{f}^{-p}+\mathcal{S}^{p}\delta\nu^{-p}
\rangle\,\nonumber\\
&=&0
\end{eqnarray}
because of $\delta a_t=0$,\,under the re-definitions of
$\tilde{\mathcal{L}}^i\,=\,3\mathcal{L}^i+5\mathcal{A}^i$ and
$\tilde{\mathcal{A}}^i\,=\,5\mathcal{L}^i+3\mathcal{A}^i$
,\,A similar calculation should be made for the barred sector.\,This solution finally ensures for that
$\alpha_1=-\frac{2 \pi}{k},\,\alpha_0=\frac{4 \pi}{k},\,\alpha_{-1}=-\frac{2 \pi}{k},\,\beta_{\frac{1}{2}}=\frac{\pi}{3k},\,\beta_{-\frac{1}{2}}=-\frac{\pi}{3k}$
,\,$\gamma_1=-\frac{2 \gamma}{k},\,\alpha_0=\frac{4 \pi}{k},\,\gamma_{-1}=-\frac{2 \pi}{k}
,\,\tau_{\frac{3}{2}}=\frac{\pi}{9k}
,\,\tau_{\frac{1}{2}}=\frac{\pi}{3k}
,\,\tau_{-\frac{1}{2}}=-\frac{\pi}{3k}
$
and
$\tau_{-\frac{3}{2}}=\frac{\pi}{9k}$
in($\ref{bouncondosp32}$).

\section{Summary and Comments}
In this work,\,a relation between $AdS_3$ and  $\mathfrak{osp}(1|2) \oplus \mathfrak{osp}(1|2)$ Chern\,-\,Simons theory was first reviewed.\,The Chern\,-\,Simons formulation of $AdS_3$ allows for a straightforward generalization to a higher\,-\,spin theory as in the bosonic cases.\,The higher\,-\,spin gauge fields have no propagating degrees of freedom, but we noted that there can be a large class of interesting non-trivial solutions.\,Specifically, $AdS_3$ in the presence of a tower of higher\,-\,spin fields up to spin $\frac{5}{2}$ is obtained by enlarging $\mathfrak{osp}(1|2) \oplus \mathfrak{osp}(1|2)$ to $\mathfrak{osp}(3|2) \oplus \mathfrak{osp}(3|2)$ for the most general $\mathcal{N}=(1,1)$ extended higher\,-\,spin supergravity theory in $AdS_3$.\,Finally,\,classical two copies of the $\mathfrak{osp}(3|2)_k$ affine algebra on the affine boundary and two copies of $\mathcal{SW}(\frac{3}{2},2)$  symmetry algebra on the superconformal boundary  as asymptotic  symmetry algebras,\,and also the chemical potentials related to source fields  appearing through the temporal components of the connection are obtained.\,Finally,\,higher\,-\,spin generalization of other class of boundary conditions,\,in particular for the Avery\,-\,Poojary\,-\,Suryanarayana gravity,\,that appeared in the literature for gravity case is also shown and checked.\,On the other hand,\,it is shown that  the Chern\,-\,Simons action which compatible with our boundary conditions leads to a finite action and  a well\,-\,defined variational principle  for the higher\,-\,spin fields.\,Therefore,\,one can think that this method provides a good laboratory for investigating the rich asymptotic structure of extended supergravity.

It could be rewarding to cast our results in the metric formulation,\,since this may help in lifting our boundary conditions to higher dimensions where  Chern\,-\,Simons formulation exists.\,Such generalizations are interesting in itself and have the potential for novel applications in holography.\,Finally,\,in our discussion of the most general boundary conditions of(super)gravity,\,we have barely scratched the boundary.\,Therefore,\,there could be numerous open questions that call for further investigations.\,For instance,\,which other boundary conditions can be obtained from a similar starting point? Or how to explain the puzzling result that the related geometries appear to have an entropy?

In conclusion,\,we consider it gratifying that the asymptotically $AdS$ story of the most general boundary conditions  initiated recently by Grumiller and Riegler to inspire  new and surprising developments even in the simple cases of three\,-\,dimensional gravity.

Our results presented in this paper can be extended in various ways.\,One possible extension is by enlarging $\mathfrak{sl}(2|1) \oplus \mathfrak{sl}(2|1)$ to $\mathfrak{sl}(3|2) \oplus \mathfrak{sl}(3|2)$ supergravity for the most general $\mathcal{N}=(2,2)$ extended higher\,-\,spin supergravity theory in $AdS_3$.\,In this context,\,one can finally emphasise here that the rather comprehensive analysis of the $\mathfrak{sl}(3|2) \oplus \mathfrak{sl}(3|2)$ theory for the Brown\,-\,Henneaux boundary conditions has been already given in Ref.\,\cite{Tan:2012xi}.\,The details of this possible extension will be examined in our forthcoming paper.

\section{Acknowledgments}
This work was supparted by Istanbul Technical University Scientific Research Projects Department(ITU BAP,\,project number:40199).


\end{document}